\documentclass[useAMS,usenatbib,A4]{mnras}
\usepackage{graphicx}
\usepackage{natbib}
\usepackage{amsmath}
\usepackage{amssymb}
\usepackage{caption}
\usepackage{subcaption}
\def\aap{A\&A~}
\def\mnras{MNRAS~}

\title[Complex lags from compact coronae]{Large and complex X-ray time lags from black hole accretion disks with compact inner coronae}
\author[P. Uttley and J. Malzac]{Phil Uttley$^{1}$\thanks{E-mail:
p.uttley@uva.nl} and Julien Malzac$^{2,3}$\\
$^{1}$Anton Pannekoek Institute, University of Amsterdam, Science Park 904, 1098 XH Amsterdam, The Netherlands\\
$^{2}$Universit\'{e} de Toulouse, UPS-OMP, IRAP, Toulouse, France\\
$^{3}$CNRS, IRAP, 9 Av. colonel Roche, BP44346, F-31028 Toulouse cedex 4, France
}

\begin{document}

\date{Submitted...}

\pagerange{\pageref{firstpage}--\pageref{lastpage}} \pubyear{2023}

\maketitle

\label{firstpage}

\begin{abstract}
Black hole X-ray binaries in their hard and hard-intermediate states display hard and soft time lags between broadband noise variations (high-energy emission lagging low-energy and vice versa), which could be used to constrain the geometry of the disk and Comptonising corona in these systems. Comptonisation and reverberation lag models, which are based on light-travel delays, can imply coronae which are very large (hundreds to thousands of gravitational radii, $R_{g}$) and in conflict with constraints from X-ray spectral modelling and polarimetry. Here we show that the observed large and complex X-ray time lags can be explained by a model where fluctuations are generated in and propagate through the blackbody-emitting disk to a relatively compact ($\sim$10~$R_{g}$) inner corona. The model naturally explains why the disk variations lead coronal variations with a Fourier-frequency dependent lag at frequencies $<1$~Hz, since longer variability time-scales originate from larger disk radii. The propagating fluctuations also modulate successively the coronal seed photons from the disk, heating of the corona via viscous dissipation and the resulting reverberation signal. The interplay of these different effects leads to the observed complex pattern of lag behaviour between disk and power-law emission and different power-law energy bands, the energy-dependence of power-spectral shape and a strong dependence of spectral-timing properties on coronal geometry. The observed spectral-timing complexity is thus a natural consequence of the response of the disk-corona system to mass-accretion fluctuations propagating through the disk.
\end{abstract}


\section{Introduction}
\label{sec:intro}
The X-ray spectra of hard state black hole X-ray binaries (BHXRBs) show two primary continuum components: low-temperature ($k_{\rm B}T<0.4$~keV) blackbody emission from the accretion disk and power-law-like emission produced by inverse Compton scattering of `seed' photons in hot ($k_{\rm B}T_{\rm e}\sim 100$~keV) thermal plasma of optical depths $\tau\sim 1$ (e.g. \citealt{Zdziarskietal1998,Makishimaetal2008}). For moderate- to high- luminosity hard states, the seed photons should primarily originate from the disk \citep{Doneetal2007}.  Observed X-ray reflection features further show that the disk reprocesses at least tens of per cent of the power-law luminosity, implying that the power-law emitting region has some vertical extent compared to the thin disk, although the exact geometry is still being debated (e.g. \citealt{Buissonetal2019}, \citealt{Zdziarskietal2021}).  Thus we can refer to this hot, Compton-scattering region as a `corona' without making any further assumptions about its geometry. 

\begin{figure}
    \includegraphics[width=0.47\textwidth]{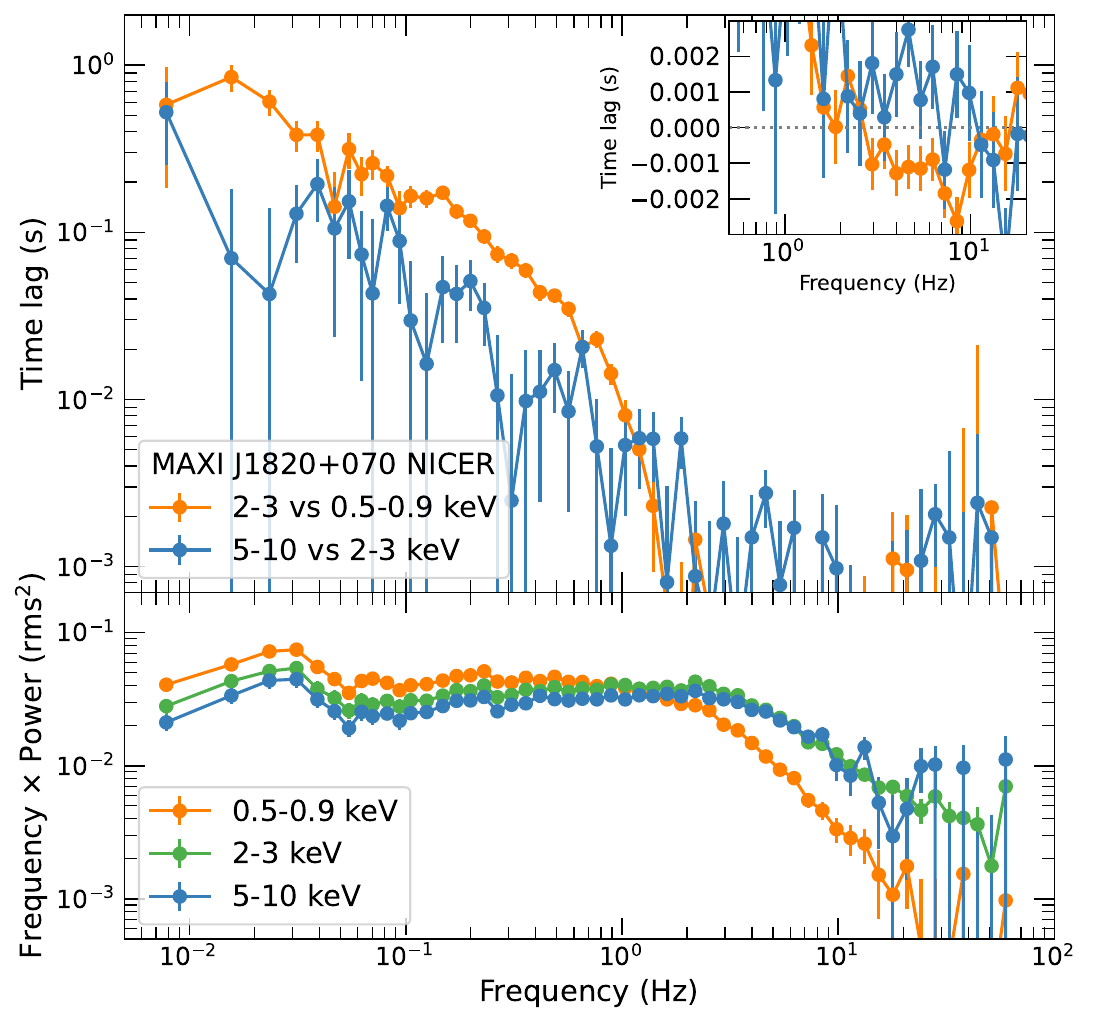}
    \caption{MAXI~J1820+070 hard state frequency-dependent lags ({\it top} panel) and power-spectra ({\it lower} panel) measured from data obtained with {\it NICER} on 2018 March 15, plotted as a function of Fourier frequency for the 0.5--0.9~keV, 2--3~keV and 5--10~keV bands.}
    \label{fig:maxij1820_lagpsd}
\end{figure}

Both disk and coronal emission in the hard state are highly variable, showing correlated aperiodic noise variability over a broad range of time-scales \citep{WilkinsonUttley2009}. Significant constraints on the relationship of the disk and corona as well as their size scale and relative geometry can be provided by X-ray spectral-timing, such as the measurement of Fourier time lags and power spectra for different energy bands. Fig.~\ref{fig:maxij1820_lagpsd} shows example lags and power spectra for disk (0.5--0.9~keV) and power-law-dominated (2--3 keV, 5--10 keV) energy bands, for hard state data obtained for the black hole X-ray binary MAXI~J1820+070\footnote{ObsID 1200120104, with 6~ks exposure obtained 2018 March 15 and processed using the approach outlined in \citet{WangJetal2022}. See \citet{Uttleyetal2014} for details on the analysis methods.} with the {\it Neutron star Interior Composition Explorer} ({\it NICER}, \citealt{Gendreauetal2016}). 

The hard lags between power-law continuum photons (henceforth, PL-PL hard lags) with variations at higher photon energies lagging those at lower energies, were the first to be discovered and studied in some detail (in Cyg~X-1, \citealt{Miyamotoetal1988,Cuietal1997,Nowaketal1999}). The lags are typically $<1$~per~cent of the variability time-scale but depend on Fourier frequency $\nu$, with time lags decreasing with increasing Fourier frequency roughly as $\tau \propto \nu^{-0.7}$, albeit with a `stepped' appearance (e.g. see \citealt{Nowak2000} and \citealt{Misraetal2017}). The lags also depend on energy $E$, with the time lag between energies $E_{2}$ and $E_{1}$, $\tau \propto \log(E_{2}/E_{1})$ \citep{Nowaketal1999,Kotovetal2001}. The evolution of PL-PL hard lags through the hard state for multiple BHXRBs shows a general trend of average lag increasing with the coronal power-law continuum photon index $\Gamma$ \citep{Pottschmidtetal2003,Grinbergetal2014,AltamiranoMendez2015,Reigetal2018}.

The power-law vs. disk lags (henceforth PL-D lags) were first discovered in {\it XMM-Newton} data for GX~339-4 \citep{Uttleyetal2011} and show apparently ubiquitous behaviour across many sources, which is notably more complex than seen for the PL-PL lags (e.g. \citealt{DeMarcoetal2015,DeMarcoetal2017,WangJetal2022}). At low frequencies the lags are positive and substantially larger than the PL-PL lags seen for the same ratio of energies at the same frequencies. At high-frequencies however the lags turn over to become negative, i.e. soft lags, of order $1-10$~ms, with the soft lag amplitude tracking the timing and spectral-state evolution closely from the hard state through to the hard to soft-intermediate state transition \citep{Karaetal2019,DeMarcoetal2021,WangJetal2021,WangJetal2022}.

Hard state power spectra also show a significant energy dependence. Besides the larger low-frequency variability amplitude at lower energies\footnote{The low-frequency energy-dependence corresponds to the short-time-scale `softer-when-brighter' behaviour of moderate--high luminosity hard state BHXRBs \citep{Skipperetal2013,SkipperMcHardy2016}}, seen in Fig.~\ref{fig:maxij1820_lagpsd}, the power-spectrum shows a steeper high-frequency slope at lower energies \citep{Nowaketal1999,Grinbergetal2014}. This effect is particularly clear for the disk band (see also \citealt{DeMarcoetalpsds2015}).

To date there is no model which can explain the complex pattern of spectral-timing behaviour of both disk and power-law emission. For example, PL-PL hard lags and power-spectral energy-dependence have been attributed to the differential light-travel delays expected from inverse-Compton scattering of photons to different energies in a spherical corona \citep{Kazanasetal1997,Huaetal1999} or jet \citep{Reigetal2003,Gianniosetal2004} of radially-varying density. Since they link the lags to the spectral-formation process, jet Comptonisation models are able to explain correlations between the continuum power-law photon index $\Gamma$ and PL-PL hard lag amplitude in terms of the jet geometric parameters \citep{Kylafisetal2008}. However, such models imply extremely vertically extended and probably also outflowing power-law emitting regions, which predict substantially flatter emissivity profiles and steeper illuminating spectra than are inferred from reflection model fits to data (e.g. see Figs. 3 and 4 in \citealt{ReigKylafis2021} for the calculated flat illumination pattern and steep illuminating spectrum). Furthermore, pure Comptonisation models do not posit a physically-motivated origin of the variability, assuming simply that seed photons 
 vary only where they are released at the base of the corona. This should lead to very short delays between the disk seed photons and lower-energy Comptonised photons, contrary to the large observed PL-D hard lags.

A solution to these difficulties can be found if we assume that variations are driven by mass-accretion fluctuations which propagate through the accretion flow, i.e. the PL-D lags are linked to the travel time of fluctuations between the disk and a central corona. Propagating accretion fluctuations are a natural feature of turbulent accretion flows \citep{Lyubarskii1997} and can explain a broad variety of aspects of XRB variability, such as their broadband power-spectral shapes \citep{Churazovetal2001,IngramDone2011}, linear rms-flux relations and log-normal flux distributions \citep{Uttleyetal2005,Heiletal2012}, the similarity between neutron star and black hole timing behaviour and its evolution \citep{SunyaevRevnivtsev2000,GardenierUttley2018} and XRB spectral-timing properties \citep{Kotovetal2001,ArevaloUttley2006}. 

To date, propagating fluctuation models have focussed on the PL-PL hard lags, by assuming that the corona corresponds to a hot inner accretion flow, i.e. within the disk truncation radius, with harder emission produced more centrally. Lags are produced by the propagation delays between radii with softer and harder emission, decreasing for smaller radii, which correspond to higher variability frequencies generated in the flow. Successively more complex versions of the model have been proposed to explain the energy and frequency-dependence of the lag and the power spectrum, by invoking radial discontinuities in: the emission spectrum  \citep{Rapisardaetal2016,MahmoudDone2018}; input variability signal \citep{MahmoudDone2018b} and speed of propagation \citep{Kawamuraetal2022}. However, these models assume ad hoc prescriptions for the radial dependence of the power-law continuum that best model the data, so they do not have explanatory power for e.g. the $\Gamma$-lag correlation. They also do not explain the PL-D lags well (or they do not explain them at all), nor do they consider the role of disk seed photons in Comptonisation. 

Recent interest in the use of X-ray spectral-timing to constrain the disk-corona geometry has focussed on X-ray reverberation, where disk reprocessing of the illuminating coronal emission produces disk emission which lags the coronal variations by the light-travel delay from corona to disk. Reverberation thus provides a natural explanation of the transition from hard to soft PL-D lags seen at high frequencies \citep{Uttleyetal2011}. The role of reverberation in producing the short-term soft lags was later confirmed with the {\it NICER} detection of delays in broadened iron K$\alpha$ reflection \citep{Karaetal2019}. Current state-of-the-art reverberation models calculate light travel delays using relativistic ray-tracing and apply them to sophisticated reflection models to fit the lags over a broad energy range \citep{Mastroserioetal2018,Mastroserioetal2021}. For simplicity, a lamppost source geometry is commonly used, but more extended coronal geometries are being investigated \citep{Lucchini2023}. However, the large soft lags push current reverberation models to infer large coronal heights (tens to hundreds of $R_{g}$, \citealt{WangJetal2021}) which, while not as extreme as predicted by pure Comptonisation models of PL-PL hard lags, remain difficult to reconcile with reflection spectral fits. Furthermore, the models do not include intrinsic disk variability and rely on fitted empirical prescriptions for power-law pivoting to account for the observed hard lags. Reconciling reverberation delays with propagation delays to model other spectral-timing behaviour remains challenging (see \citealt{Mahmoudetal2019} for a recent example).

In summary then, individual lag models fall well short of a complete explanation of the complex spectral-timing behaviour observed in black hole X-ray binaries in the hard state. To some extent, Comptonisation, propagation and reverberation must all play a role in solving the problem, but current implementations of these effects are limited because they neglect the other aspects, e.g. either the physical origin of the continuum spectrum (in the case of propagation or reverberation models) or the origin of variability itself (in the case of Comptonisation models). 

In this paper, we show how to resolve these difficulties by implementing a model which includes the effects of propagation and reverberation together with a self-consistent approximation for power-law continuum formation in the corona. We do so by making the simplifying assumption that, on the observed variability time-scales, the Comptonising corona is in thermal equilibrium with respect to seed and coronal heating variations, with normalisation and $\Gamma$ set by the seed photon flux and the ratio of seed photon to coronal heating luminosities (as expected from conservation of both photon number and total power in the corona). 
\begin{figure*}
    \includegraphics[width=0.95\textwidth]{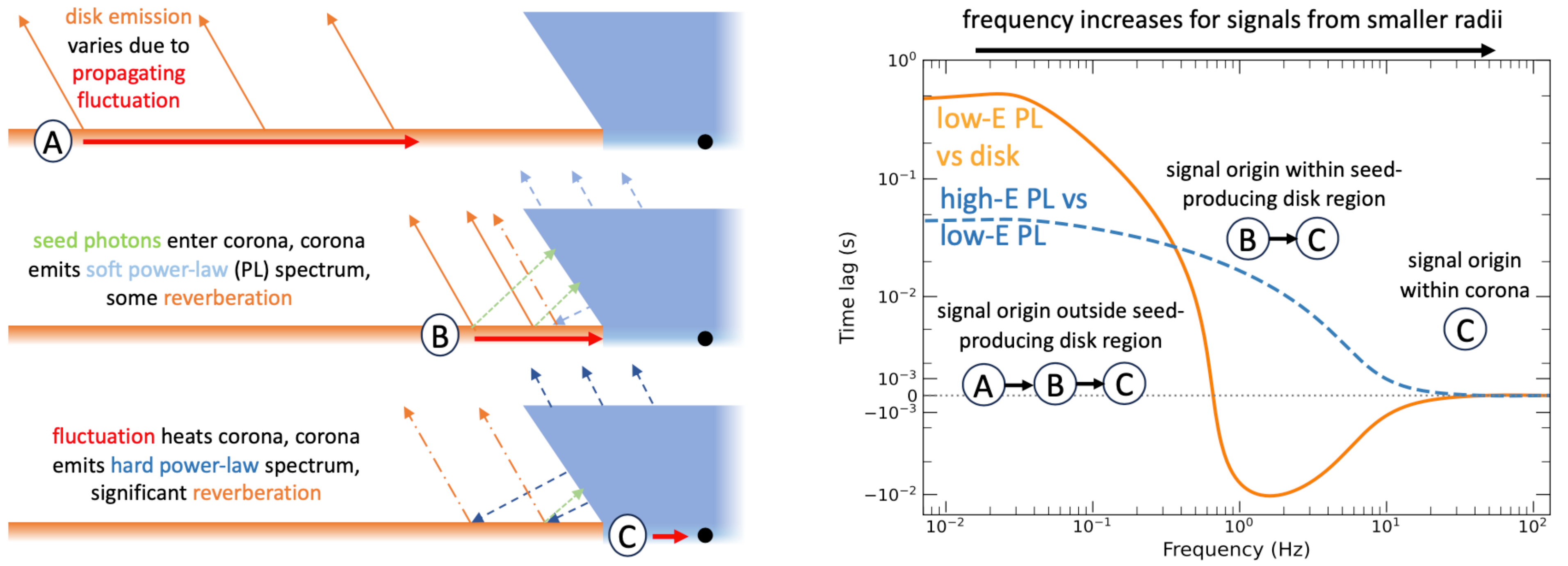}
    \caption{Summary of the propagating fluctuation lag model presented in this work. A fluctuation signal will produce a combination of the effects $A$, $B$ and $C$ (shown {\it left}) depending on the radius of origin of the signal. Since the radius of origin determines the signal frequency, frequency-dependent lags are seen ({\it right}, plotted schematically with $\sinh^{-1}$ scaling to show positive and negative lags on the same axis). For signals originating at large radii ($A$) the large propagation delays dominate the lags, leading to positive power-law vs. disk lags as $A$ is followed by $B$ and $C$. However, for signals produced within the seed emitting region, the disk emission is dominated by the reverberation component (dot-dashed arrows) which is produced mostly in $C$ and lags the soft power-law emission produced mostly at $B$ by the propagation delay between $B$ and $C$. Frequency-dependent hard lags between power-law bands are also produced due to the delay between $B$ and $C$. For the highest frequencies, which are produced in the corona, propagation-related delays are negligible and light-travel delays may start to dominate (not shown in the calculated lags).}
    \label{fig:lags_diagram}
\end{figure*}

Our model is shown schematically in Fig.~\ref{fig:lags_diagram}. Following previous propagation models, mass accretion fluctuations are generated with radially-dependent time-scales in both the disk and corona, which we treat as an inner hot flow but with a vertical extent described by the coronal geometry. Fluctuations propagating through the disk naturally produce the observed frequency-dependent PL-D hard lags and suppression of soft band variability at high frequencies. Seed photons are produced in the disk and are modulated by accretion fluctuations before those same fluctuations reach the corona and modulate the coronal heating rate. This means that the coronal power-law spectrum pivots in response to any given fluctuation, with steepening in response to slower fluctuations from further out in the disk and hardening in response to variations propagating through the corona. This effect naturally produces hard PL-PL lags on long time-scales as well as flatter high-frequency power spectra at harder power-law continuum energies. 

Although our model was originally conceived to simultaneously explain only the PL-D and PL-PL hard lags, the inclusion of reverberation of coronal luminosity on the disk naturally produces large soft PL-D lags at high Fourier frequencies, even for compact ($\sim 10$~$R_{g}$) coronae and under the assumption of negligible light-travel delays. These lags arise from the delay between disk reverberation, which is driven by the total coronal power and the soft power-law emission, which is driven by seed photon variations and thus precedes the peak in coronal power due to the propagation delay of fluctuations between the inner disk and corona. The model can naturally explain changes in PL-PL hard lags and PL-D soft lags, in terms of changes in the visibility of disk seed photons from the corona that are linked to changes in coronal geometry. 

Here in this work, we introduce the model in its most basic form, considering the lags between disk total luminosities and mono-energetic power-law bands, and treating the seed photons as being mono-energetic. We consider the dependence of lags on disk photon energy briefly in the Appendix, but will carry out a full energy-dependent treatment and fits to data in later papers. In Section~\ref{sec:irfmodel} we use an impulse response function approach to show that hard power-law lags with a log-linear energy dependence are automatically produced by the corona when seed photons vary before coronal heating.   In Section~\ref{sec:var_acc_flow} we show how this effect is automatically realised by a variable accretion flow and we describe a numerical implementation of this model in Section~\ref{sec:num_imp}. In Section~\ref{sec:results} we show how the model reproduces most of the key spectral-timing results discussed so far.  We discuss the model implications and its limitations in Section~\ref{sec:discussion}, before setting out our conclusions in Section~\ref{sec:conclusions}.  In the Appendix we demonstrate some key results and use numerical Monte Carlo simulations of variable accretion to demonstrate the validity of our impulse response approach in the non-linear variability regime.

\section{Time response of coronal emission}
\label{sec:irfmodel}
We consider Comptonising coronae in thermal equilibrium, where electrons are heated via some unspecified heating mechanism with heating rate given by a luminosity $L_{\rm h}$. This assumption is justfied for coronae with light-crossing times $<100$~$R_{g}/c$ and the expected moderate optical depths ($\tau \sim 1$), and is observationally supported by the high coherence of variations in different energy bands in the hard state, which imply that we see temporally connected emitting regions, with any delays described by simple linear transforms \citep{Nowaketal1999}. The electrons are cooled via inverse Compton scattering of seed photons with luminosity $L_{\rm s}$. We then assume that the instantaneous photon index $\Gamma$ is set by the ratio of the time-dependent input seed luminosity to the coronal heating luminosity as follows:
\begin{equation}
\label{eqn:gameqn}
\Gamma(t) = \Gamma_{0}\left(\frac{L_{\rm s}(t)}{L_{\rm h}(t)}\right)^{\beta} \quad ,
\end{equation}
where $\Gamma_{0}$ is the photon index for equal seed and heating luminosities and $\beta$ is an index which may be derived from the specific Comptonisation model (e.g. see \citealt{PietriniKrolik1995} and \citealt{Beloborodov2001}). 

Since the model is linearised, the precise form of the relation between photon index and cooling/heating ratio does not matter as much as the direction of the relation (which sets the sign of the lags). A positive correlation between $\Gamma$ and the seed-to-heating luminosity ratio, arises naturally from conservation of both total luminosity and photon number in the corona. The assumption of thermal equilibrium is valid when the heating and cooling time-scales within the corona are small compared to the time-scales of variability of seed and heating luminosities. This should be the case for the relatively compact coronae considered here.

To calculate the variation of a quantity $f(t)$ in response to a perturbing signal $s(t)$, we require the impulse response function (henceforth `impulse response') for that quantity $g(\tau)$ (where $\tau$ is the time delay), so that the perturbation in $f(t)$, $\delta f(t) = \int^{\infty}_{-\infty} s(t-\tau) \ast g(\tau) {\rm d}\tau$. We will now derive the impulse responses for power-law continuum slope and flux variations, in terms of the coronal seed and heating variations.

For our simplified analytical calculation, we make the approximation that the seed photons are monochromatic with energy $E_{\rm s}$, so that the total photon number is $L_{\rm s}/E_{\rm s}$.  Assuming that total photon numbers are conserved during the Comptonisation process and that the cut-off photon energy $E_{\it cut} \gg E_{\rm s}$, the photon flux density variation at an energy $E$, $N(E,t)$ is given by:
\begin{equation}
\label{eqn:plfluxeqn}
N(E,t)= \frac{L_{\rm s}(t)(\Gamma(t)-1)}{E_{\rm s}^{2}}\left(\frac{E}{E_{\rm s}}\right)^{-\Gamma(t)}\quad .
\end{equation}
By linearising Equations~\ref{eqn:gameqn} and \ref{eqn:plfluxeqn}, i.e. assuming small perturbations in the time-variable parameters we obtain the impulse responses for $\Gamma(t)$ and $N(E,t)$:
\begin{equation}
\label{eqn:gamirf}
g_{\Gamma}(\tau) = \langle \Gamma \rangle \beta \left(\frac{g_{\rm s}(\tau)}{\langle L_{\rm s} \rangle} - \frac{g_{\rm h}(\tau)}{\langle L_{\rm h} \rangle} \right)
\end{equation}
\begin{equation}
\label{eqn:fluxirf}
g_{\rm pl}(E,\tau) = \langle N(E) \rangle \left([1-u(E)]\frac{g_{\rm s}(\tau)}{\langle L_{\rm s} \rangle} + u(E) \frac{g_{\rm h}(\tau)}{\langle L_{\rm h} \rangle}\right)
\end{equation}
where $u(E)=\beta \langle \Gamma \rangle \left(\ln (E/E_{\rm s}) - \frac{1}{\langle \Gamma \rangle-1}\right)$ and angle brackets denote the time-averaged quantities which are perturbed by changes in the seed and heating luminosities. Those changes are themselves quantified in terms of their own impulse responses, $g_{\rm s}$ and $g_{\rm h}$, which will depend on the physical system producing the seed and heating luminosities and their variations. We will determine these functions for the case of propagating accretion fluctuations in the next Section.

The $u(E)$ term contains the energy-dependence of the impulse response shape, as it weights the heating and seed impulse responses differently according to the energy of power-law photons being considered. Effectively, $u(E)$ produces pivoting of the power-law spectrum about an energy $E_{\rm piv}$, which corresponds to $g_{\rm pl}(E_{\rm piv},\tau)=0$ and is given by:
\begin{equation}
\label{eqn:plpivot}
    \ln \left(\frac{E_{\rm piv}}{E_{\rm s}}\right)=\left[\beta\langle \Gamma \rangle\left(1-\frac{g_{\rm h}(\tau)}{\langle L_{\rm h}\rangle}\big/ \frac{g_{\rm s}(\tau)}{\langle L_{\rm s} \rangle}\right)\right]^{-1}+\frac{1}{\langle \Gamma \rangle -1}
\end{equation}

For $\beta=1/6$ \citep{Beloborodov2001} and typical observed photon indices $1.4 < \langle \Gamma \rangle < 2.5$, if a change in seed luminosity dominates, the {\it minimum} pivot energy is $E_{\rm piv}\approx21$--$885 E_{\rm s}$ (with the smallest pivot energy for the steepest spectra), which corresponds to $g_{\rm h}(\tau)=0$. As $g_{\rm h}(\tau)$ increases relative to  $g_{\rm s}(\tau)$, the pivot energy increases rapidly until $g_{\rm h}(\tau)/\langle L_{\rm h}\rangle=g_{\rm s}(\tau)/\langle L_{\rm s}\rangle$, when $E_{\rm piv}=\infty$ and only the power-law normalisation changes, while photon index does not vary. If $g_{\rm h}(\tau)/g_{\rm s}(\tau)$ continues to increase, the pivot switches to very low energies, increasing towards a limiting {\it maximum} energy $E_{\rm piv}\approx2$--$12 E_{\rm s}$ (with steeper spectra corresponding to lower maximum pivot energies), which corresponds to $g_{\rm s}=0$. Thus, differences between the time responses of the seed and heating luminosities lead to changes in the spectral pivot-point and significant time-delay dependent changes in where the largest and smallest power-law flux variations are seen. This mechanism naturally leads to energy-dependent lags in the power-law emission, as well as energy dependence of the power-law flux variability amplitude.

\section{Impulse responses from a variable accretion flow}
\label{sec:var_acc_flow}
We assume that the driving signals are mass accretion rate fluctuations propagating through the disk.  For simplicity we will assume simple non-diffusive propagation, so that the amplitudes and power-spectral shapes of mass-accretion fluctuations are preserved as they propagate inwards.  This approximation can be applied to `classical' accretion disks provided that the fluctuation time-scale is comparable to or greater than the viscous time-scale at the radius it is generated \citep{Churazovetal2001,Mushtukovetal2018}.

The response delay $\tau$\footnote{Throughout this paper we use $\tau$ to denote a relative time delay, while $t$ is used to represent a time or time-scale.} depends on the radial drift velocity of the accretion variations.  Since the final model will combine signals from multiple radii, it is useful to set $\tau=0$ to be the time when a propagating signal reaches the innermost radius of the accretion flow, $r_{\rm in}$, which allows us to define a fixed, negative, propagation time-delay for each radius $r$ (where $r$ is expressed in units of the gravitational radius $R_{g}$):
\begin{equation}
\label{eqn:tau}
\tau(r) =  \int^{r}_{r_{\rm in}} {\rm d}\tau = \int^{r}_{r_{\rm in}} {\rm d}r/v_{r}
\end{equation}
where the radial drift velocity $v_{r} = -\alpha (h/r)^{2} r^{-1/2}$, for a disk with scale-height $h$ and dimensionless viscosity parameter $\alpha$.  Note that $v_{r}$ is expressed as a fraction of the speed-of-light $c$, so the delays are naturally expressed in units of the light-crossing time for 1~$R_{g}$.  

For simplicity we assume that light-travel time lags, e.g from the disk to the corona, are negligible compared to the lags due to propagating fluctuations which we consider here. We have also assumed that $r_{\rm in}$ is the inner zero-torque boundary of the whole accretion flow (including the coronal part), which we assume to correspond to the innermost stable circular orbit (ISCO). It's also important to note that the radial dependence of propagation delays (and time-scales of variability, discussed later) will be significantly affected by the radial scaling of the disk aspect ratio $h/r$.

We define all luminosities as fractions of the total dissipated power from the flow.  The power dissipated by the accretion flow in an annulus ${\rm d}r$ at radius $r$ is therefore defined, using the usual radial dependence (e.g. \citealt{FKR}), as:
\begin{equation}
\label{eqn:f_diss}
f_{\rm diss}(r){\rm d}r=\frac{r^{-2}[1-(r_{\rm in}/r)^{1/2}]{\rm d}r}{\int^{r_{\rm out}}_{r_{\rm in}} r^{-2} [1-(r_{\rm in}/r)^{1/2}]{\rm d}r}
\end{equation}
where $r_{\rm out}$ is the outer radius of the flow.  At this stage we speak in terms of the accretion flow rather than the accretion disk, because we assume that the central compact corona is itself powered by accretion and may be thought of as a hot inner flow with radius $r_{\rm cor}$, which corresponds to the truncation radius of the blackbody-emitting disk.  

\subsection{Coronal heating and seed luminosity}
\label{sec:coronal_irfs}
If we assume that all power dissipated within $r_{\rm cor}$ goes into heating the corona, we can define the coronal heating impulse response function $g_{\rm h}(\tau)$ as:
\begin{equation}
g_{\rm h}(\tau){\rm d}\tau =
\begin{cases}
f_{\rm diss}(r){\rm d}r, & \text{if } r_{\rm in} < r \leq r_{\rm cor} \\
0, & \text{otherwise}
\end{cases}
\end{equation}
Seed photons are provided by the disk. If the fraction of disk emission from a radius $r$ which is intercepted by the corona is $f_{\rm d\rightarrow c}(r)$, the seed photon impulse response due to disk photons produced by viscous dissipation is:
\begin{equation}
\label{eqn:irf_seeddisp}
g_{\rm s, diss}(\tau){\rm d}\tau =
\begin{cases}
f_{\rm d\rightarrow c}(r) f_{\rm diss}(r){\rm d}r, & \text{if } r_{\rm cor} < r \leq r_{\rm fluc} \\
0, & \text{otherwise}
\end{cases}
\end{equation}
where $r_{\rm fluc}$ is the starting radius of the accretion fluctuation (which we assume to begin in the disk, i.e. outside the corona).  Note that we include the additional subscript `diss' to make a distinction between disk (and seed) emission which is produced directly by dissipation in the flow, and that produced by reprocessing of the coronal power-law emission by the disk. The calculation of $f_{\rm d\rightarrow c}(r)$ depends on the assumed coronal geometry and is discussed in Appendix~\ref{app:geomcalc}. For the present work, we assume that the disk is the only source of seed photons to the corona. It would be simple however, to allow some fraction of the viscous dissipation powering the corona to be converted into internal seed photons, e.g. via synchrotron emission if the corona is magnetised \citep{Veledinaetal2013}.

A fraction of the emitted total coronal luminosity (which is the sum of seed and heating luminosities) will be intercepted and reprocessed by the disk to produce an X-ray reverberation signal which has a short, light-travel time-delay with respect to the coronal X-ray variations. Some fraction of this reverberation component will return to the corona and can provide an extra source of seed photons. The fraction, $f_{\rm return}$, of the total coronal luminosity which returns to the corona in this way is:
\begin{equation}
f_{\rm return}=2\pi\int^{r_{\rm out}}_{r_{\rm cor}} f_{\rm d\rightarrow c}(r) f_{\rm c\rightarrow d}(r)\,r\,{\rm d}r
\end{equation}
where $f_{\rm c\rightarrow d}(r)$ is the fraction per unit area of coronal photons intercepted by the disk, as a function of disk radius (see Appendix~\ref{app:geomcalc} for details of its calculation). 

The returning photons will be inverse Compton-scattered in the corona and may undergo additional reprocessing in the disk to contribute higher orders of returning luminosity. Assuming negligible light-travel delays, to maintain coronal energy balance we obtain for the time-dependent total returning luminosity $L_{\rm return}(t)$:
\begin{equation}
\label{eqn:returnlum}
    \begin{split}
        L_{\rm return}(t) & = f_{\rm return}\left(L_{\rm h}(t)+L_{\rm s, diss}(t)+L_{\rm return}(t)\right) \\
        & = f_{\rm return} \left(\frac{L_{\rm h}(t)+L_{\rm s, diss}(t)}{1-f_{\rm return}}\right)
    \end{split}
\end{equation}
A few tens of per cent of the returning luminosity is in the form of backscattered, predominantly harder X-rays (the so-called `reflection hump'). The remaining luminosity, besides a few per cent in isolated fluorescent emission lines such as Fe K$\alpha$, may be thermalised by the disk, or comprise of softer reprocessed emission, which for higher disk densities common in X-ray binaries can start to approximate the smooth blackbody continuum shape of the disk thermal emission \citep{RossFabian2007,Garciaetal2016,Mastroserioetal2021}. For simplicity, we assume that this latter `thermal reverberation' component provides returning seed photons to the corona at similar energies to the seed photons from viscous dissipation in the disk, with luminosity equal to a fraction $f_{\rm therm}$ of the total returning luminosity.  We implicitly assume that the remaining returning photons either do not affect the coronal spectral shape or may heat or cool the corona in a way which is subsumed by $f_{\rm therm}$, which acts as a kind of efficiency factor for the contribution of returning luminosity towards coronal cooling. Using this factor and Eqn.~\ref{eqn:returnlum}, we infer the impulse response of the seed luminosity associated with thermal reverberation:
\begin{equation}
\label{eqn:irf_seedrev}
g_{\rm s, rev}(\tau){\rm d}\tau = \frac{f_{\rm therm}f_{\rm return}}{(1-f_{\rm return})}\left[g_{\rm s, diss}(\tau) + g_{\rm h}(\tau)\right]{\rm d}\tau
\end{equation} 

\subsection{Direct disk emission and thermal reverberation}
\label{sec:disk_irfs}
The direct disk emission is the non-reprocessed disk blackbody emission which is due to viscous dissipation, observed at infinity (averaging over all observer viewing angles).
\begin{equation}
\label{eqn:irf_diskdisp}
g_{\rm d, diss}(\tau){\rm d}\tau =
\begin{cases}
(1-f_{\rm d\rightarrow c}(r)) f_{\rm diss}(r){\rm d}r, & \text{if } r_{\rm cor} < r \leq r_{\rm fluc} \\
0, & \text{otherwise}
\end{cases}
\end{equation}
To conserve luminosity, the direct disk emission is reduced by the fraction of the disk emission which is intercepted by the corona, although we note that for a complete treatment the angular dependence of this component should be accounted for. We must also account for the part of the disk reverberation signal considered in the previous section which goes directly to the observer instead of returning to the corona. This component will track the total coronal luminosity variation with negligible delay in our propagating fluctuations model. The total fraction of coronal luminosity that is intercepted and reprocessed by the disk, $f_{\rm rev}$, is equal to:
\begin{equation}
f_{\rm rev} = 2\pi \int^{r_{\rm out}}_{r_{\rm cor}} f_{\rm c\rightarrow d}(r)\,r\,{\rm d}r \:,
\end{equation}
The reverberation signal is enhanced by the contribution of the returning luminosity to the coronal emission and must also be corrected for it (since the returning luminosity is not observed directly as reverberation). We also only consider the (quasi-)thermal part of the reverberation signal, which contributes to the total impulse response of disk thermal emission. The resulting impulse response of disk thermal reverberation is:
\begin{equation}
\label{eqn:irf_diskrev}
g_{\rm d, rev}(\tau){\rm d}\tau = \frac{f_{\rm therm}(f_{\rm rev}-f_{\rm return})}{(1-f_{\rm return})}\left[g_{\rm s, diss}(\tau) + g_{\rm h}(\tau)\right]{\rm d}\tau
\end{equation}

\subsection{Total impulse responses, luminosities and power-law flux impulse response}
\label{sec:total_irfs}
The final total disk and seed photon impulse responses are:
\begin{equation}
g_{\rm d}(\tau){\rm d}\tau = \left[g_{\rm d, diss}(\tau)+g_{\rm d, rev}(\tau)\right]{\rm d}\tau
\end{equation}
\begin{equation}
g_{\rm s}(\tau){\rm d}\tau = \left[g_{\rm s, diss}(\tau)+g_{\rm s, rev}(\tau)\right]{\rm d}\tau
\end{equation}
The time-averaged luminosities for the heating, disk and seed components can be determined by integrating the impulse responses over all radii (implicitly setting $r_{\rm fluc}=r_{\rm out}$ for the purposes of this calculation) e.g. $\langle L_{\rm h} \rangle = \int^{\tau(r_{\rm out})}_{\tau(r_{\rm in})} g_{\rm h}(\tau){\rm d}\tau$. Note that using the impulse response definitions described in Sections~\ref{sec:coronal_irfs} and \ref{sec:disk_irfs} the total luminosity reaching the observer is correctly conserved, i.e.:
\begin{multline} 
\langle L_{\rm total} \rangle = \langle L_{\rm d, diss} \rangle + \langle L_{\rm d, rev} \rangle/f_{\rm therm}
+ \\ \left(1-f_{\rm rev}\right)\left(\langle L_{\rm h}\rangle + \langle L_{\rm s, diss} \rangle + \langle L_{\rm s, rev} \rangle/f_{\rm therm}\right) = 1
\end{multline}
where the factors including $f_{\rm therm}$ and $f_{\rm rev}$ correct respectively for the non-thermal part of the reflection spectrum and the fraction of coronal luminosity reaching the observer. 

Finally, the impulse response of power-law emission can be calculated from the impulse responses and time-averaged luminosities for the coronal heating and total seed component derived above, using Eqn.~\ref{eqn:fluxirf}. The time-averaged photon index $\langle \Gamma \rangle$ is part of the pivoting term $u(E)$ and can be determined by using the seed and heating time-averaged luminosities in Eqn.~\ref{eqn:gameqn}, which is appropriate for the linear approximation of power-law variability used in this work. Note that the time-averaged power-law flux can be calculated from the time-averaged quantities using Equation~
\ref{eqn:plfluxeqn}, although since it is only a normalising factor for the flux impulse response it does not have any influence on our results.

\subsection{Calculation of spectral-timing products}
\label{sec:st_products}
The impulse responses derived above can be used to determine spectral-timing properties in response to a mass-accretion fluctuation signal which originates at a single radius $r_{\rm fluc}$.  In reality it is likely that propagating fluctuations originate over a range of radii, as surmised by the model of \citet{Lyubarskii1997} and supported by the broadband nature of the observed power spectra and multi-time-scale rms-flux relation observations (e.g. see discussion in \citealt{Uttleyetal2005}) as well as (general relativistic) magnetohydrodynamic simulations of accretion flows \citep{HoggReynolds2016,Bollimpallietal2020}.  Since the simple impulse response functions described above map delay times on to unique radii, it is simple to model multi-location signals by using the same impulse responses but with different starting delays.  Given a starting radius $r_{\rm fluc}$ we can use Eqn.~\ref{eqn:tau} to define a starting delay time $\tau_{\rm fluc}$, which is negative since we define the innermost delay to be zero.  The impulse response $g_{\rm fluc}$ `seen' by that fluctuating signal is then related to the overall impulse response ($g(\tau)$) calculated for that component by:
\begin{equation}
\label{eqn:sigirf}
g_{\rm fluc}(\tau) =
\begin{cases}
g(\tau) & \text{if } \tau \geq \tau_{\rm fluc} \\
0, & \text{otherwise}
\end{cases}
\end{equation}
If we then define $n_{\rm fluc}$ signals with Fourier power-spectra $P_{{\rm fluc},i}(\nu)$ (where $\nu$ is the Fourier frequency), originating at radii $r_{{\rm fluc}, i}$ with corresponding time delays $\tau_{{\rm fluc},i}$, the resulting power-spectrum is given by:
\begin{equation}
\label{eqn:powspec}
P(\nu) = \sum_{i=1}^{n_{\rm fluc}} P_{{\rm fluc},i}(\nu) |G_{{\rm fluc},i}(\nu)|^{2}
\end{equation}
where $G_{{\rm fluc},i}(\nu)$ is the Fourier transform of the impulse response for $\tau_{{\rm fluc},i}$ given by Eqn.~\ref{eqn:sigirf}.  The cross-spectrum between two components, labelled 1 and 2 (e.g. two energies in the power-law, or the disk vs. a single power-law energy) is:
\begin{equation}
\label{eqn:crossspec}
C_{1,2}(\nu) =  \sum_{i=1}^{n_{\rm fluc}} P_{{\rm fluc},i}(\nu) G_{{\rm fluc},i,1}(\nu)G_{{\rm fluc},i,2}^{\ast}(\nu)
\end{equation}
with frequency-dependent time-lag given by the argument of the cross-spectrum (the phase lag) normalised by $2\pi\nu$.

For simplicity we assume that the power spectra of the individual mass-accretion rate fluctuation signals $P_{{\rm fluc},i}(\nu)$, are described by Lorentzian functions parameterised by a peak frequency $\nu_{{\rm fluc},i}$, quality-factor $Q_{i}$ (approximately the frequency of the Lorentzian divided by its full-width at half maximum) and fractional rms $rms_{i}$. As discussed by \citet{Pottschmidtetal2003}, the more common form of the Lorentzian function can then be calculated from these parameters, by determining the resonance frequency $\nu_{{\rm res}, i}=\nu_{{\rm fluc}, i}/\sqrt{1+1/(4Q_{i}^{2})}$ and normalising factor $R_{i}=rms_{i}/\sqrt{0.5-\tan^{-1}(-2Q_{i})/\pi}$:
\begin{equation}
    P_{{\rm fluc},i}(\nu) = \frac{2R_{i}^{2}Q_{i} \nu_{{\rm res}, i}}{\pi\left(\nu_{{\rm res}, i}^{2}+4Q_{i}^{2}(\nu-\nu_{{\rm res}, i})^{2}\right)}
\end{equation}
The number and radial locations of Lorentzian signals and their parameters can be selected when setting up the model calculation and in principle can be arbitrary. However, the linear model described here does not account for viscous diffusion effects on mass accretion fluctuations as they propagate through the flow, which significantly suppresses variations on time-scales exceeding the local viscous propagation time-scale (e.g. see discussion in \citealt{Churazovetal2001} and a more extensive treatment in \citealt{Mushtukovetal2018}). Therefore for simplicity, we will require that signal peak frequencies only correspond to time-scales which exceed the propagation delay through the flow from that radius.

The above equations and assumed input signal power spectra allow us to calculate the observed power spectra and lags expected from our model. It is important to note that this approach to combining signals from different radii assumes that the mass accretion rate variations combine additively instead of the multiplicative combination implicit in the propagating fluctuations model \citep{Uttleyetal2005}.  We will consider the effects of this simplification in Appendix~\ref{app:nonlin}, by comparing our linear impulse response model predictions with predictions from numerical simulations of multiplicative propagating fluctuations. 

\section{Numerical implementation}
\label{sec:num_imp}

To calculate the disk, seed and heating impulse responses, we define 200 contiguous, geometrically spaced radial bins of radii $r_{i}$ and widths $\Delta r_{i}$ between $r_{\rm in}=2$~$R_{g}$ (corresponding to an ISCO with dimensionless spin parameter $a=0.94$) and $r_{\rm fluc, max}$, which corresponds to the outermost radius where mass-accretion fluctuations are generated in the disk. To account for the constant emission from the non-variable part of the disk, we further define a set of 200 contiguous, geometrically spaced disk radii from $r_{\rm fluc, max}$ to $r_{\rm out}=1000$~$R_{g}$ which is used to calculate the constant parts of disk and seed emission, which are included in the mean luminosities of those components, in addition to the contributions calculated from the impulse responses within $r_{\rm fluc, max}$. For calculating the reverberation component of disk emission, we set $f_{\rm therm}=0.7$, corresponding approximately to 1 minus the disk albedo for X-ray reflection in the absence of soft X-ray reprocessing.

\subsection{Mass accretion fluctuation time-scales and power-spectrum}
\label{sec:mdotfluctuations}
We assume that mass accretion fluctuations are generated in both the disk and corona in each radial bin within $r_{\rm fluc, max}$, i.e. there are $n_{\rm fluc}=200$ separate fluctuation signals. The fluctuation signals (see Section~\ref{sec:st_products}) have Lorentzian power spectra with peak frequencies $\nu_{{\rm fluc},i}=\left((R_{g}/c)t_{{\rm fluc},i}\right)^{-1}$~Hz and constant quality-factor $Q_{i}=1$. The fluctuation time-scale $t_{{\rm fluc},i}$ is given by:
\begin{equation}
\label{eqn:timescales}
t_{{\rm fluc},i} =
\begin{cases}
s_{t}r_{i}^{(n_{t,{\rm d}}+\frac{3}{2})}, & \text{if } r_{\rm fluc, max} \geq r_{i} > r_{\rm cor} \\
s_{t}f_{t,{\rm c}} r_{\rm cor}^{(n_{t,{\rm d}}-n_{t,{\rm c}})} r_{i}^{(n_{t,{\rm c}}+\frac{3}{2})}, & \text{if } r_{\rm cor} \geq r_{i} > r_{\rm in}
\end{cases}
\end{equation}
where $s_{t}$ is a normalising constant and $n_{t,{\rm d}}$, $n_{t,{\rm c}}$ are indices which correspond to the radial scaling of the fluctuation time-scale with respect to the local dynamical time-scale in the disk and corona respectively. The parameter $f_{t,{\rm c}}$ is a scaling factor between time-scales of variability in the disk and corona. For an accretion flow described by viscosity parameter $\alpha$ \citep{ShakuraSunyaev1973}, $s_{t}r_{i}^{n_{t}} \propto \alpha^{-1}(r/h)^{2}$, so that (assuming constant $\alpha$) a radially-constant aspect ratio $h/r$ corresponds to $n_{t}=0$, while constant scale-height $h$ corresponds to $n_{t}=2$. 

To obtain a reasonable match to observations, we fix $s_{t}=1$ and $n_{t,{\rm d}}=2$, corresponding to a disk scale-height which is constant with radius. The main argument for a steep radial dependence of disk variability time-scale, i.e. $n_{t,{\rm d}}=2$, is that it better matches the observed energy-dependent low-frequency PL-D hard lags, which we demonstrate in Appendix~\ref{app:diskencalc} but do not consider further here, except to note that the spectral-timing properties discussed for $n_{t,{\rm d}}=2$ in Section~\ref{sec:results} show similar properties for $n_{t,{\rm d}}=0$ or 1. We assume $f_{t,{\rm c}}=1$ which corresponds to the disk and corona fluctuation time-scales being equal at $r_{\rm cor}$. We assume that the coronal time-scale scaling index is given by $n_{t, {\rm c}} = \max\{0, n_{t,{\rm c}}(t_{\rm min})\}$, where $t_{\rm min}$ is a time-scale which must equal or exceed the time-scale of variability fluctuations generated at $r_{\rm in}$, from which $n_{t,{\rm c}}(t_{\rm min})$ can be calculated by rearranging the lower equation in Equation~\ref{eqn:timescales}. With this constraint, we limit the coronal scaling index $n_{t,{\rm c}}\geq 0$ (i.e. the limiting case is a flow with constant aspect ratio), and the fastest variability time-scale produced in the flow to be $t_{{\rm fluc},1}\lesssim t_{\rm min}$.

The shapes of broadband BHXRB power spectra (e.g. \citealt{Heiletalpowcol2015}) lead us to assume a natural form for the mass accretion fluctuations which is a doubly-bending power-law with $P(\nu)\propto \nu^{-1}$ in the intermediate frequency range (i.e. flat in $\nu P(\nu)$). This shape can be approximately reproduced by weighting the squared fractional rms of each Lorentzian by the logarithmic change in peak frequency to the next radial bin\footnote{For this purpose, we also include the frequency calculated using Equation~\ref{eqn:timescales} for the non-fluctuating bin at $i=n_{\rm fluc}+1$.}:
\begin{equation}
\label{eqn:rnsvals}
rms_{i}^{2}=rms_{\rm tot}^{2}\frac{\log(\nu_{{\rm fluc},i}/\nu_{{\rm fluc},i+1})}{\log(\nu_{{\rm fluc},1}/\nu_{{\rm fluc},n_{\rm fluc}+1})}
\end{equation}
where $rms_{\rm tot}$ is the total fractional rms of the mass accretion fluctuation signals, which we assume to be 40\%. The magnitude of rms is important for considering non-linear effects, both due to power-law photon index variations and the multiplicative nature of mass accretion variations implied by the rms-flux relation. However, in Appendix~\ref{app:nonlin} we use numerical simulations to show that these effects remain negligible for accretion fluctuations with integrated rms of 40\%, consistent with the X-ray variability observed in hard state BHXRBs.

We determine the bend-frequencies of the mass-accretion fluctuation power-spectrum by setting the maximum and minimum fluctuation time-scales such that the frequency at the maximum fluctuation radius $r_{\rm fluc,max}$ is $\nu_{{\rm fluc,max}}=\left((R_{g}/c)t_{{\rm fluc},i=n_{\rm fluc}}\right)^{-1}$~Hz and at the minimum fluctuation radius (terminating at $r_{\rm in}$), $\nu_{{\rm fluc,min}}=\left((R_{g}/c)t_{{\rm fluc},i=1}\right)^{-1}$~Hz. For the calculations shown here we choose $\nu_{{\rm fluc,max}}=0.03$~Hz (corresponding to $r_{\rm fluc,max}=46$~$R_{g}$ for $s_{t}=1$, $n_{t,{\rm d}}=2$), and set $t_{\rm min}$ to correspond to a frequency of 100~Hz, which sets the maximum fluctuation frequency to equal or exceed this value. Together, these choices will result in power-spectra similar to those seen in BHXRBs for at least part of the hard state.  
\begin{figure*}
    \centering
    \begin{subfigure}[t]{0.54\textwidth}
        \centering
        \includegraphics[width=\linewidth]{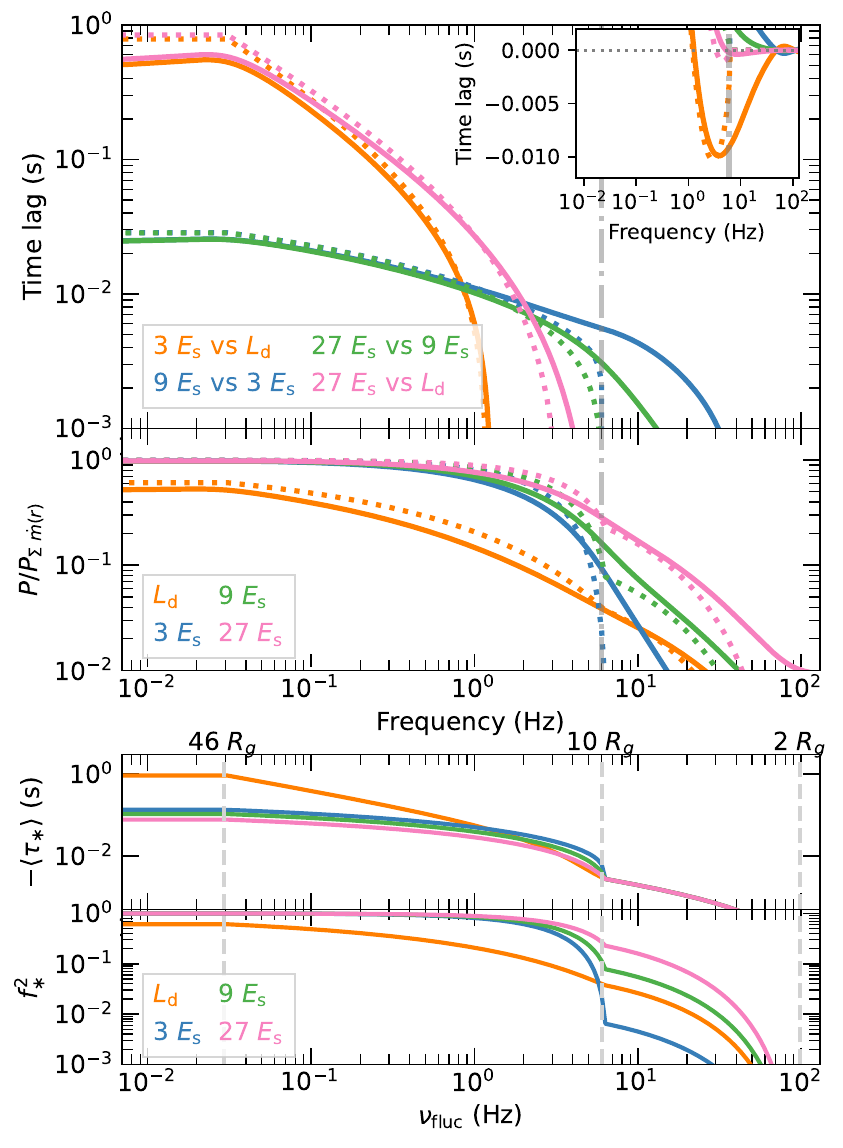} 
    \end{subfigure}
    \begin{subfigure}[t]{0.44\textwidth}
        \centering
        \includegraphics[width=\linewidth]{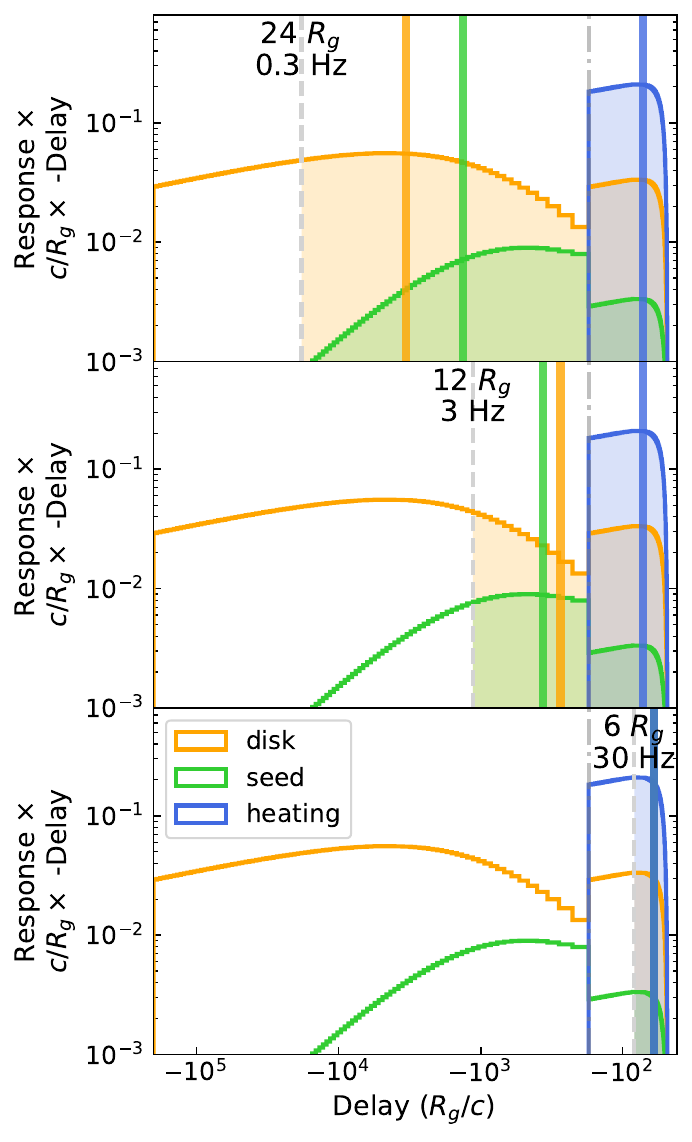} 
    \end{subfigure}
    \caption{Model spectral-timing properties and their relation to impulse response functions, for a spherical corona with radius $r_{\rm cor}=10$~$R_{g}$. {\it Top-left}: Calculated frequency-dependent time-lag and power-spectral ratio to input fluctuation power spectrum, for the disk luminosity and several power-law energies. The solid lines show the full model calculations, while dotted lines show an approximation based on the impulse response centroid delays and emission fractions vs. fluctuation signal frequency (see text for details). The frequency-dependences of these quantities are shown in the {\it lower-left} panel. The {\it right} panel shows the disk, seed and coronal heating impulse responses for this configuration, showing from top to bottom the part of the impulse responses (shaded) which responds to a signal at a given radius and corresponding frequency and delay (vertical dashed lines). The centroid delay values for each of the starting signal radii are shown as solid vertical lines in the colour corresponding to the component (disk, seed and heating). The delays are plotted using symmetric logarithmic axes with delays $>-100$~$R_{g}/c$ plotted on a linear scale. The response units are luminosity per unit time delay, where luminosities are expressed as a fraction of the total dissipated accretion luminosity. The impulse responses are further multiplied by the absolute time delay, so that the delays which contribute the most to the impulse response appear as a peak.} \label{fig:irf_explain}
\end{figure*}

\subsection{Propagation delays}
We assume that the propagation delay across a given radial bin $\Delta\tau_{i}$, scales with the local variability time-scale in that bin. This behaviour is expected if fluctuation time-scales scale with the local viscous time-scale, as expected in the propagating fluctuations scenario \citep{Lyubarskii1997,Churazovetal2001,HoggReynolds2016,Mushtukovetal2018}. Based on Equation~\ref{eqn:tau} and incorporating additional scaling factors for propagation in the disk and corona, $f_{\Delta\tau,{\rm d}}$ and $f_{\Delta\tau,{\rm c}}$, we obtain the following relations for the propagation delay across a radial bin:
\begin{equation}
\label{eqn:delays}
\Delta\tau_{i} =
\begin{cases}
s_{t}f_{\Delta\tau,{\rm d}}r_{i}^{(n_{t,{\rm d}}+\frac{1}{2})}\Delta r_{i}, & \text{if } r_{\rm fluc, max} \geq r_{i} > r_{\rm cor} \\
s_{t}f_{\Delta\tau,{\rm c}} r_{\rm cor}^{(n_{t,{\rm d}}-n_{t,{\rm c}})} r_{i}^{(n_{t,{\rm c}}+\frac{1}{2})}\Delta r_{i}, & \text{if } r_{\rm cor} \geq r_{i} > r_{\rm in}
\end{cases}
\end{equation}

The resulting delays are used to calculate the impulse response functions following the approach outlined in Sections~\ref{sec:coronal_irfs} and \ref{sec:disk_irfs}. Our default assumption is that $f_{\Delta\tau,{\rm d}}=1$ $f_{\Delta\tau,{\rm c}}=0.1$. With these scalings and those of the disk and corona variability time-scales, we expect a smooth evolution of variability time-scale from disk to corona, but a transition to a fast accretion time-scale within the corona, commensurate with the expected geometrically thick and optically thin hot flow.

\subsection{Spectral-timing calculations}
To enable rapid calculation of the Fourier transforms of impulse responses required for determining spectral-timing products, we approximate our impulse response functions to be sums of shifted top-hat functions, which correspond to each radial delay bin. I.e., following Equation~\ref{eqn:sigirf}, the continuous-time impulse response applied to fluctuations produced in the $j$th radial bin is:
\begin{equation}
    g_{{\rm fluc},j}(\tau) = \sum_{k=1}^{j} g(\langle \tau\rangle_{k}) \mbox{rect}(\tau/\Delta\tau_{k}) \ast \delta(\tau-\langle\tau\rangle_{k})
\end{equation}
where $g(\langle\tau\rangle_{k})$ is the impulse response value for the bin with mean delay $\langle \tau\rangle_{k} =\left(\sum_{l=1}^{k} \Delta\tau_{l}\right)-\Delta\tau_{k}/2$, $\mbox{rect}(\tau/\Delta\tau_{k})$ is the normalised rectangular function centred on zero with total width $\Delta\tau_{k}$ and the product of these quantities is then time-shifted to $\tau=\langle\tau\rangle_{k}$ by convolution with a $\delta$-function. The Fourier transform of a time-shifted top-hat function is a phase-shifted sinc-function, so the Fourier transform of the impulse response for a signal originating in the $j$th radial bin can be calculated for any given frequency $\nu$, using:
\begin{equation}
    G_{{\rm fluc},j}(\nu) = \sum_{k=1}^{j} g(\langle\tau\rangle_{k})\Delta\tau_{k} \mbox{sinc}(\Delta\tau_{k}\nu) \exp(-i2\pi \nu \langle\tau\rangle_{k})
\end{equation}

where $i$ here denotes the unit imaginary number. With this approach, we can use Equations~\ref{eqn:powspec} and \ref{eqn:crossspec} to rapidly calculate the frequency-dependent spectral-timing properties for our chosen frequencies. We use a geometrically-spaced grid of 500 frequencies over the range 0.003--200~Hz.

\section{Results}
\label{sec:results}
We calculate the power-law flux impulse responses for soft, medium and hard photon energies $E_{\rm soft}=3E_{\rm s}$, $E_{\rm med}=9E_{\rm s}$ and $E_{\rm hard}=27E_{\rm s}$. The power-law lags for energies defined in this way are independent of the choice of seed energy $E_{\rm s}$. However, for a seed energy corresponding to the peak of a disk blackbody with maximum temperature $kT_{\rm max}$, $E_{\rm s}=2.82kT_{\rm max}$ so that the relevant energies for a hard state disk with $kT_{\rm max}=0.2$~keV are $E_{\rm s}\simeq 0.56$~keV, $E_{\rm soft}\simeq 1.7$~keV, $E_{\rm med}\simeq 5.1$~keV and $E_{\rm hard}\simeq 15.2$~keV. To compare the power spectra, we normalise them by the summed mass accretion rate signal power spectrum, $P_{\sum \dot{m}(r)}=\sum^{n_{\rm fluc}}_{i=1} P_{{\rm fluc},i}(\nu)$. By normalising in this way, we remove some of the effects of the somewhat arbitrary choice of driving signals which dominate the power-spectral shape, while maintaining the energy-dependent effects which are a key outcome of the model.

\subsection{Spectral-timing properties for a disk and spherical corona with $r_{\rm cor}=10$~$R_{g}$}
\label{sec:results_explainer}
To demonstrate and explain the X-ray spectral-timing properties predicted by our model, we first consider a spherical corona with radius $r_{\rm cor}=10$~$R_{g}$, with default fluctuation signal and propagation delay parameters given in Section~\ref{sec:num_imp}. This coronal geometry predicts a mean photon index $\langle \Gamma \rangle=1.47$, consistent with that seen in hard state black hole X-ray binaries. The frequency-dependent lags and power-spectra for different power-law energies (in terms of $E_{\rm s}$) and the disk luminosity $L_{\rm d}$ are shown by the solid lines in the upper-left panels of Fig.~\ref{fig:irf_explain}. The model shows a strong correspondence to a number of the key spectral-timing properties of hard state BHXRBs which we described in Section~\ref{sec:intro} :
\begin{enumerate}
    \item At low frequencies, disk variations substantially lead power-law variations at all energies considered showing hard PL-D lags, but cross to large soft PL-D lags above $\sim$1 Hz, when measured with respect to the lower energy power-law photons. The high-frequency disk lags are much weaker relative to higher energy power-law photons, approaching zero then becoming positive (power-law lagging) as the power-law energy increases.
    \item Power-law photons of different energies show frequency-dependent hard PL-PL lags, with a weaker and less complex frequency dependence than seen for PL-D lags.
    \item Lags between power-law energies scale equal to or close to log-linearly with energy, over a broad frequency range (i.e. the same factor change in energy shows the same lag).
    \item Variability at high-frequencies is suppressed and the amount of suppression is energy-dependent, with lower-energy power-law emission showing less high-frequency power than higher energy emission. Disk emission variability is strongly suppressed down to lower frequencies than the power-law emission.
\end{enumerate}
To understand how these spectral-timing properties arise, consider the disk, seed and coronal heating impulse responses that are `seen' by signals originating at different radii ($r_{\rm fluc}$) in the accretion flow, with fluctuation frequency $\nu_{\rm fluc}$ (see Section~\ref{sec:st_products}). These are illustrated in the right panel of Fig.~\ref{fig:irf_explain} by shaded regions, which extend to the right of the delay time $\tau_{\rm fluc}$ at the signal radius of origin, marked by vertical dashed lines. The impulse responses produced outside the coronal radius (illustrated by the vertical dot-dashed line) are generated primarily by viscous dissipation in the disk, with a weak reverberation component due to disk seed luminosity returning from the corona to heat the disk. The impulse responses produced inside the coronal radius are powered by viscous dissipation in the corona, which produces coronal heating and in turn leads to a pure-reverberation component of disk and seed emission.
\begin{figure*}
    \centering
    \begin{subfigure}[t]{0.49\textwidth}
        \centering
        \includegraphics[width=\linewidth]{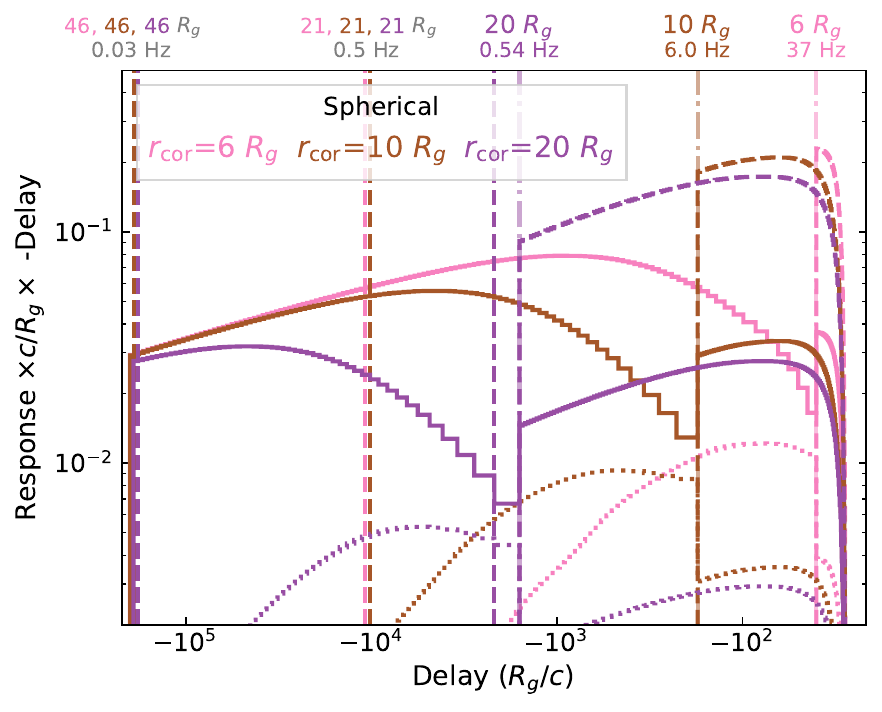} 
    \end{subfigure}
    \hfill
    \begin{subfigure}[t]{0.49\textwidth}
        \centering
        \includegraphics[width=\linewidth]{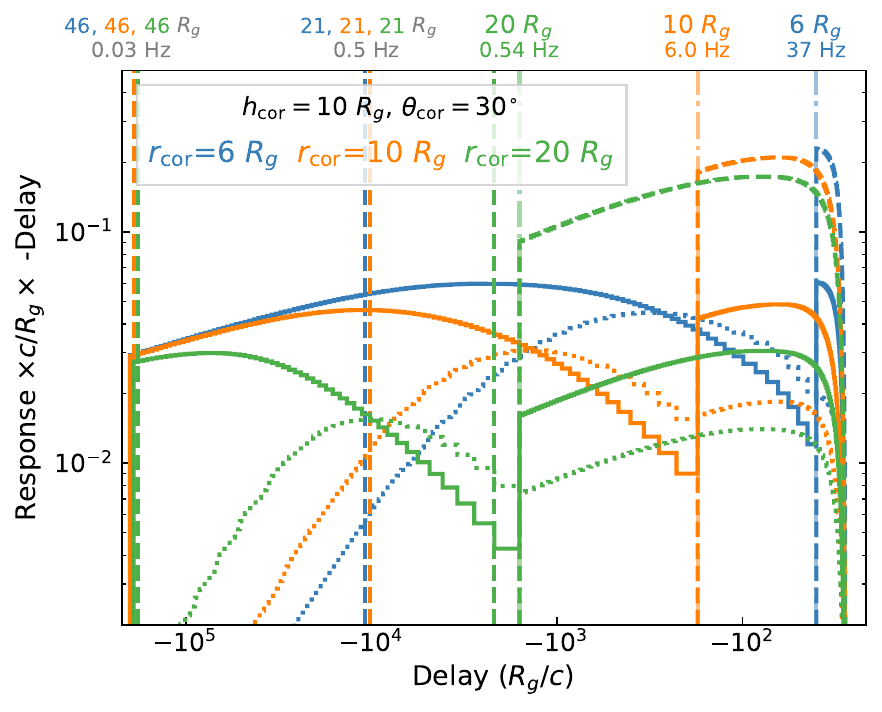} 
    \end{subfigure}
    \vspace{0.3cm}
    \begin{subfigure}[t]{0.49\textwidth}
        \centering
        \includegraphics[width=\linewidth]{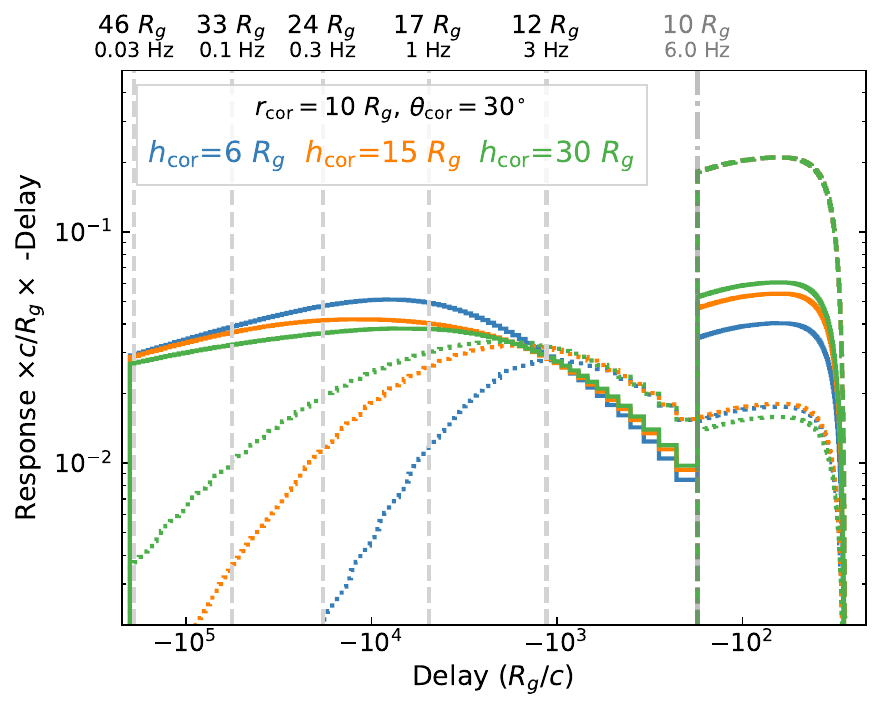} 
    \end{subfigure}
    \hfill
    \begin{subfigure}[t]{0.49\textwidth}
        \centering
        \includegraphics[width=\linewidth]{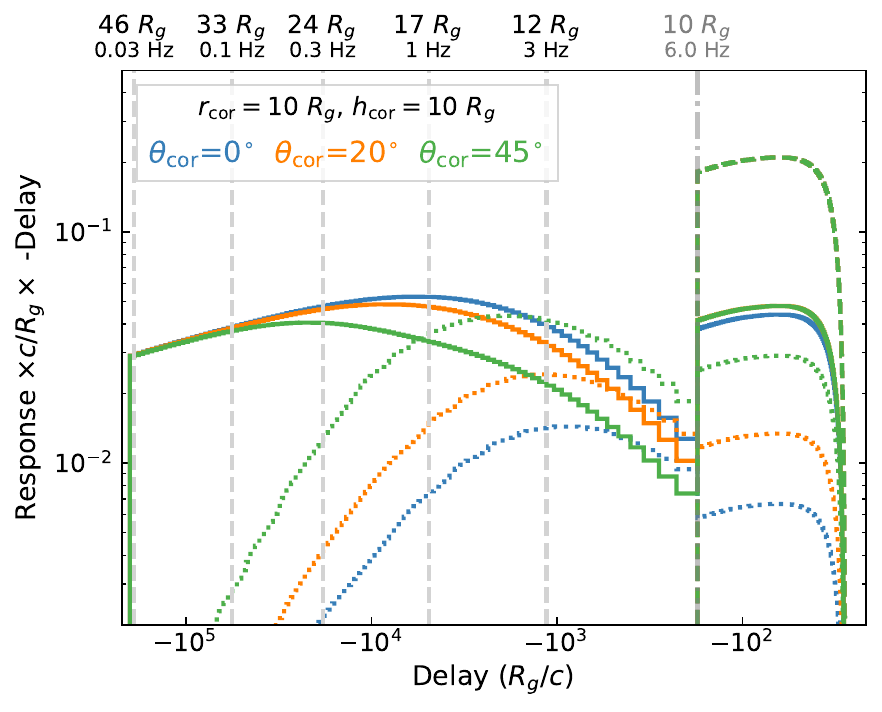} 
    \end{subfigure}
    \caption{Comparison of impulse responses for different coronal geometries and parameter variations. {\it Top-left}: changing coronal radius $r_{\rm cor}$ for a spherical corona. Then clockwise from {\it top-right}: inverted cone corona with changing coronal radius $r_{\rm cor}$; changing opening angle $\theta_{\rm cor}$; changing coronal height $h_{\rm cor}$. For each changing parameter value the plots show the total disk (solid lines), seed (dotted lines) and heating (dashed lines) impulse responses. Vertical dashed lines show the radii and fluctuation signal frequencies associated with a given delay. For the top panels with varying $r_{\rm cor}$, these radii correspond to different delays and are colour-coded accordingly. See Fig.~\ref{fig:irf_explain} for further details.} \label{fig:irfs}
\end{figure*}

To better understand how these impulse responses lead to the complex spectral-timing behaviour predicted by the model, it is useful to make some simplifying approximations. For signals with variability time-scales which are long compared to the emission-weighted (centroid) delay $\langle \tau \rangle$, as is generally the case here, the time-lag between two emission components (denoted 1 and 2), $\tau_{1,2}(\nu) \simeq \langle\tau_{2}\rangle - \langle\tau_{1}\rangle$. The centroid delay is calculated using:
\begin{equation}
    \langle \tau_{\ast} \rangle = \frac{\int^{\tau_{\rm fluc}}_{0} \tau g_{\ast}(\tau){\rm d{\tau}}}{\int^{\tau_{\rm fluc}}_{0} g_{\ast}(\tau){\rm d{\tau}}}
    \label{eqn:centroid}
\end{equation}
where $\ast$ symbolises the component, i.e. d(isk), s(eed) and h(eating). The seed and heating components are not directly observed but contribute to the power-law emission, to give the power-law emission centroid delay:
\begin{equation}
    \langle \tau_{\rm pl}(E)\rangle = \frac{[1-u(E)]f_{\rm s}\langle \tau_{\rm s}\rangle + u(E)f_{\rm h}\langle \tau_{\rm h}\rangle}{[1-u(E)]f_{\rm s} + u(E)f_{\rm h}} 
    \label{eqn:pl_centroid}
\end{equation}
where $f_{s}$, $f_{\rm h}$ correspond to the `emission fraction' of the total luminosity for that impulse response component, i.e. equivalent to the shaded areas in the right panels of Fig.~\ref{fig:irf_explain}:
\begin{equation}
    f_{\ast} = \frac{\int^{\tau_{\rm fluc}}_{0} g_{\ast}(\tau){\rm d{\tau}}}{\langle L_{\ast}\rangle}
\end{equation}
with power-law emission fractions given by $f_{\rm pl}(E)=[1-u(E)]f_{\rm s} + u(E)f_{\rm h}$.
For time-scales which are long compared to the centroid delay, the observed variability power of a fluctuation signal scales with the intrinsic power multiplied by $f_{\ast}^{2}$ or $\left(f_{\rm pl}(E)\right)^{2}$.

The numerical values of the disk and power-law centroid delays and the squared emission fractions are shown as a function of fluctuation frequency in the lower-left panel of Fig.~\ref{fig:irf_explain}. To compare our approximate predictions for lag and power-spectral amplitude with the full model calculations, we plot the fluctuation frequency-dependent centroid delay differences and squared emission fractions as dotted lines in the upper-left panel. For our simplified calculations we do not combine signals with overlapping power-spectra and simply assume that a signal dominates at its respective fluctuation frequency. Despite this approximation, the simplified calculation results are quite close to those obtained with the full Fourier treatment of combined signals.

For more extended impulse responses such as for the disk emission, the lags and relative power-spectral amplitude both decrease strongly with increasing fluctuation frequency due to the lag centroid (relative to the power-law emission) and emission fraction changes. The power-law emission on the other hand is a weighted combination of two more centrally concentrated impulse response components (seed and heating), leading to smaller lags and weaker evolution of lagas and power-spectra with fluctuation frequency, except at higher frequencies when the fluctuation signals are fully embedded in the regions producing seed and heating components. The energy-dependent weighting of the seed and heating components leads to more extended power-law impulse responses at lower energies where the seed emission contributes more strongly than heating, so that suppression of the high-frequency power-spectrum decreases with photon energy, as observed. For lower fluctuation frequencies the PL-PL lag amplitude scales linearly with the difference in $\log(E)$, also as observed (note that the lags are identical for the same ratio of energies considered). This effect arises as $f_{\rm s}$, $f_{\rm h}\rightarrow 1$ for larger fluctuation signal radii. In that limit, the lag for power-law energies $E_{1}$ and $E_{2}$ (calculated using the energy-dependent centroid delays obtained with Equation~\ref{eqn:pl_centroid}) simplifies to:
\begin{equation}
    \langle \tau(E_{2}) \rangle - \langle \tau(E_{1}) \rangle \simeq \beta \langle \Gamma \rangle \left(\langle \tau_{\rm h} \rangle - \langle \tau_{\rm s} \rangle \right)
    \ln(E_{2}/E_{1})\; .
\label{eqn:pllags_centroid}
\end{equation}
Therefore, the power-law emission lags scale linearly with the difference between heating and seed centroid delays (thus producing the observed positive, i.e. hard lags), the mean photon index, and the difference in log-$E$. The latter dependence arises naturally from the power-law pivoting predicted by the model.

Our model also reproduces the negative PL-D lags that have previously been attributed to reverberation light-travel delays \citep{Uttleyetal2011}. The PL-D negative lags arise primarily because the illumination of the disk by the corona produces a strong peak of disk reverberation emission at the smallest delays, which is not replicated for the seed emission, since only a small fraction of disk reverberation photons return to the corona. The combination of these effects causes the disk centroid for small signal radii to become smaller (less negative) than the seed and soft power-law emission centroids, so that the power-law vs. disk lag becomes negative at higher Fourier frequencies, as observed. This relative shift in the disk and seed centroid delays can be seen in the top-right and middle-right panels of Fig.~\ref{fig:irf_explain}, as the fluctuation signal shifts from 0.3 Hz to 3 Hz.

These `soft' lags are primarily due to the reverberation component of disk emission, but they are much larger than any expected light travel delay, as they are linked to the propagation of fluctuations through the inner disk radii which provide most of the seed photons. The disk lags disappear when comparing with higher energy power-law emission, also as observed, because of the increased effect of heating of the corona at those energies, which has an impulse response concentrated at small delays, similar to the reverberation component which is also driven by the coronal heating luminosity.

\subsection{Effects of coronal geometry on impulse responses and spectral-timing properties}
\label{sec:results_corgeom}
The coronal geometry has a significant impact on the predicted lags, for several reasons. Firstly, the coronal visibility from the disk determines the amount of seed emission intercepted from different disk radii. A more vertically extended corona will intercept proportionately more photons from fluctuations further out in the disk, to produce a seed impulse response that extends to larger negative delays. Secondly, coronae that are more visible from the disk will illuminate the disk more strongly to produce stronger reverberation components in the disk and seed impulse response at small delays. The coronal radius also determines the distribution of accretion power between disk emission (and hence seed luminosity) and coronal heating, which affects the mean photon index as well as the relative strength of reverberation vs. propagation components in the disk and seed impulse responses.

To examine the effect of coronal geometry on spectral-timing properties, we will consider two types of coronal geometry: spherical (with radius $r_{\rm cor}$) and an inverted cone geometry. The inverted cone shape (strictly speaking an inverted {\it conical frustum}, i.e. a cone with the pointed end cut off) is intended to represent a jet or wind-like corona or a cylindrically-shaped `hot flow'. It has a basal coronal radius in the disk plane $r_{\rm cor}$ and coronal height above the disk plane $h_{\rm cor}$. The radius at the top (widest) end of the cone is determined by the opening angle $\theta_{\rm cor}$ which is the angle relative to the normal to the disk plane, such that $\theta_{\rm cor}=0^{\circ}$ gives a cylindrical corona. Unlike the spherical corona, the inverted cone geometry allows us to study the effect of changes in coronal visibility as seen from the disk, independent of changes in coronal radius.

\begin{figure*}
    \includegraphics[width=0.98\textwidth]{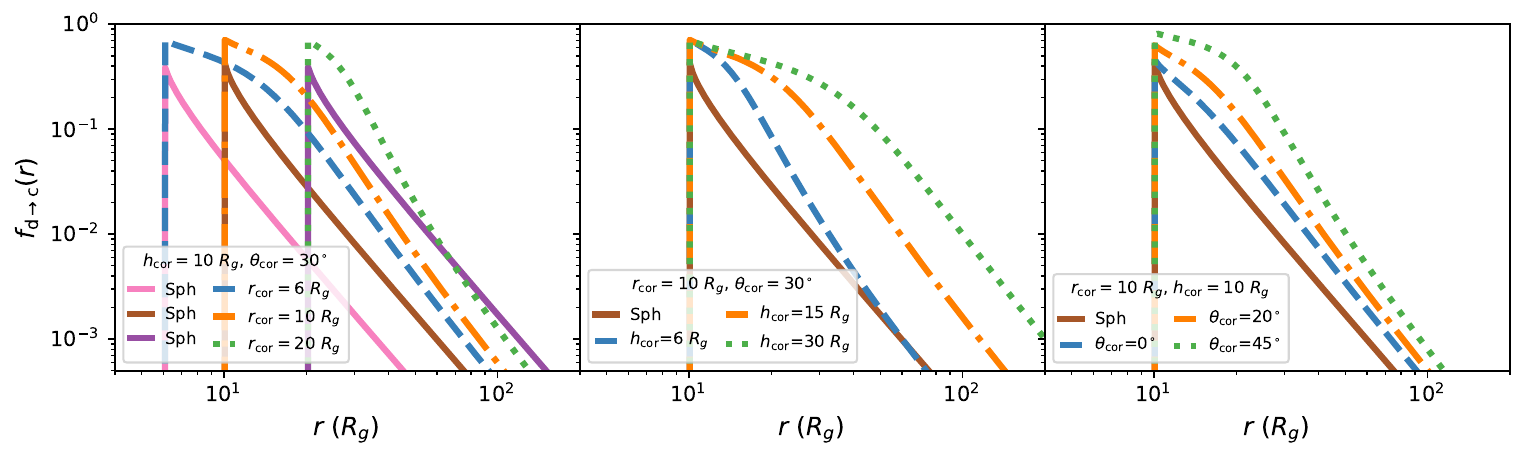}
    \caption{Comparison of the radially-dependent fraction of disk photons intercepting the corona $f_{\rm d\rightarrow c}(r)$, for a spherical corona (solid lines) and an inverted cone corona (dashed, dot-dashed and dotted lines) with different radii (left panel), heights (middle panel) and opening angles (right panel).}
    \label{fig:fd2c}
\end{figure*}

We consider the effects of changing the coronal radius (for both the spherical and inverted-cone geometries), and the coronal height and opening angle (for the inverted-cone geometry). Figure~\ref{fig:irfs} shows the disk (solid lines), seed (dotted lines) and heating (dashed lines) impulse responses for these situations, with both the dissipated and reverberation contributions to the disk and seed components summed to give those impulse responses. To understand the differences in impulse responses between the different geometries, it will be useful to consider the visibility of the corona from the disk for the different geometries, which we can show by plotting the radially-dependent fraction of disk photons intercepted by the corona, $f_{\rm d\rightarrow c}(r)$, in Fig.~\ref{fig:fd2c}. 

\begin{figure*}
    \centering
    \begin{subfigure}[t]{0.49\textwidth}
        \centering
        \includegraphics[width=\linewidth]{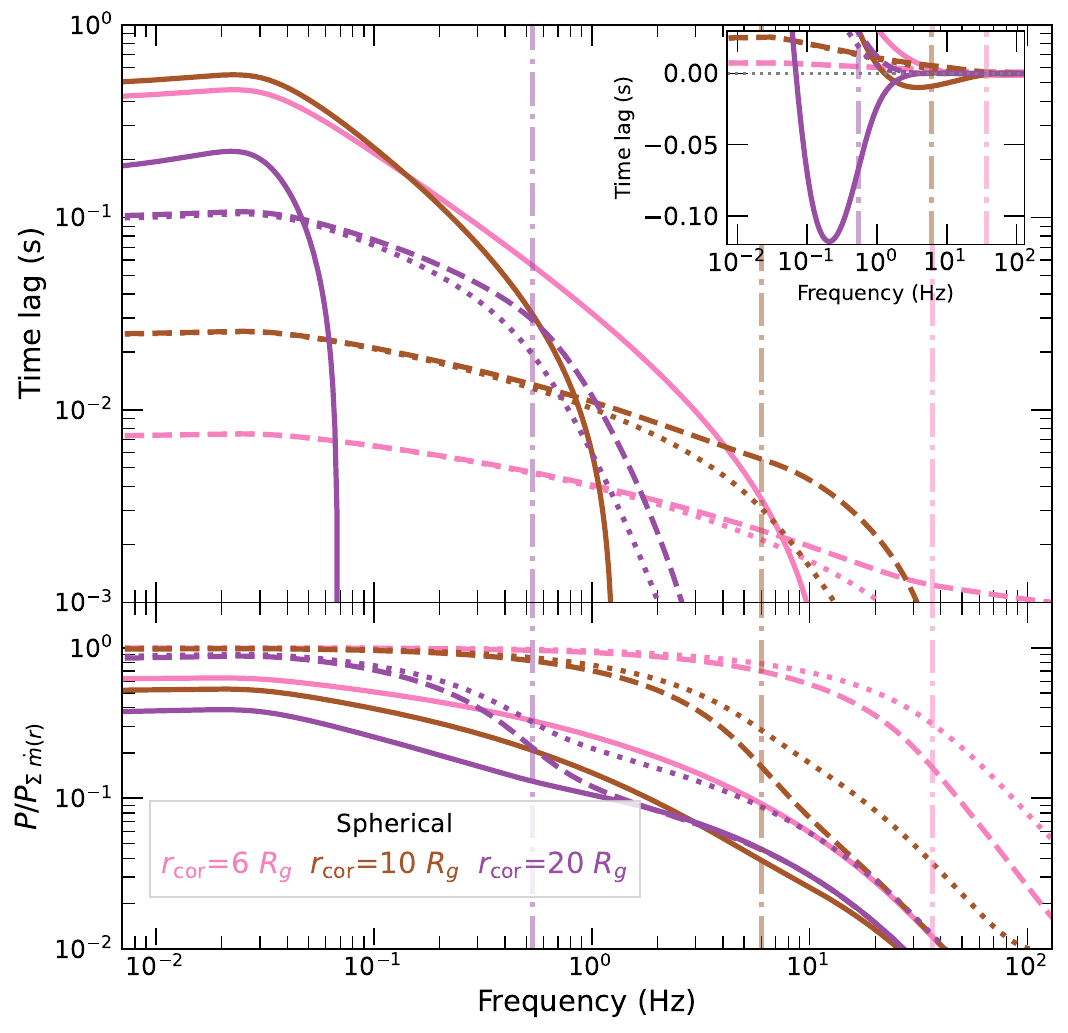} 
    \end{subfigure}
    \hfill
    \begin{subfigure}[t]{0.49\textwidth}
        \centering
        \includegraphics[width=\linewidth]{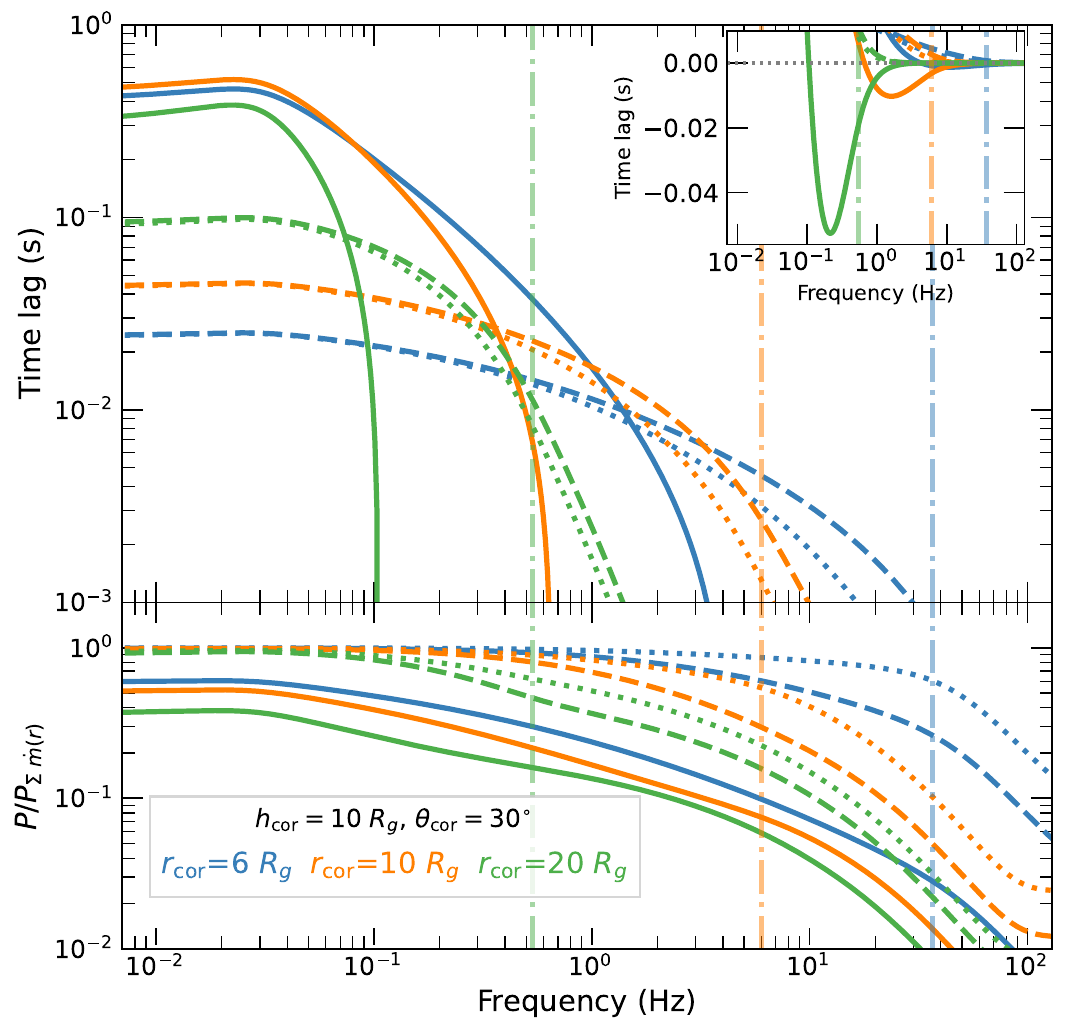} 
    \end{subfigure}
    \vspace{0.3cm}
    \begin{subfigure}[t]{0.49\textwidth}
        \centering
        \includegraphics[width=\linewidth]{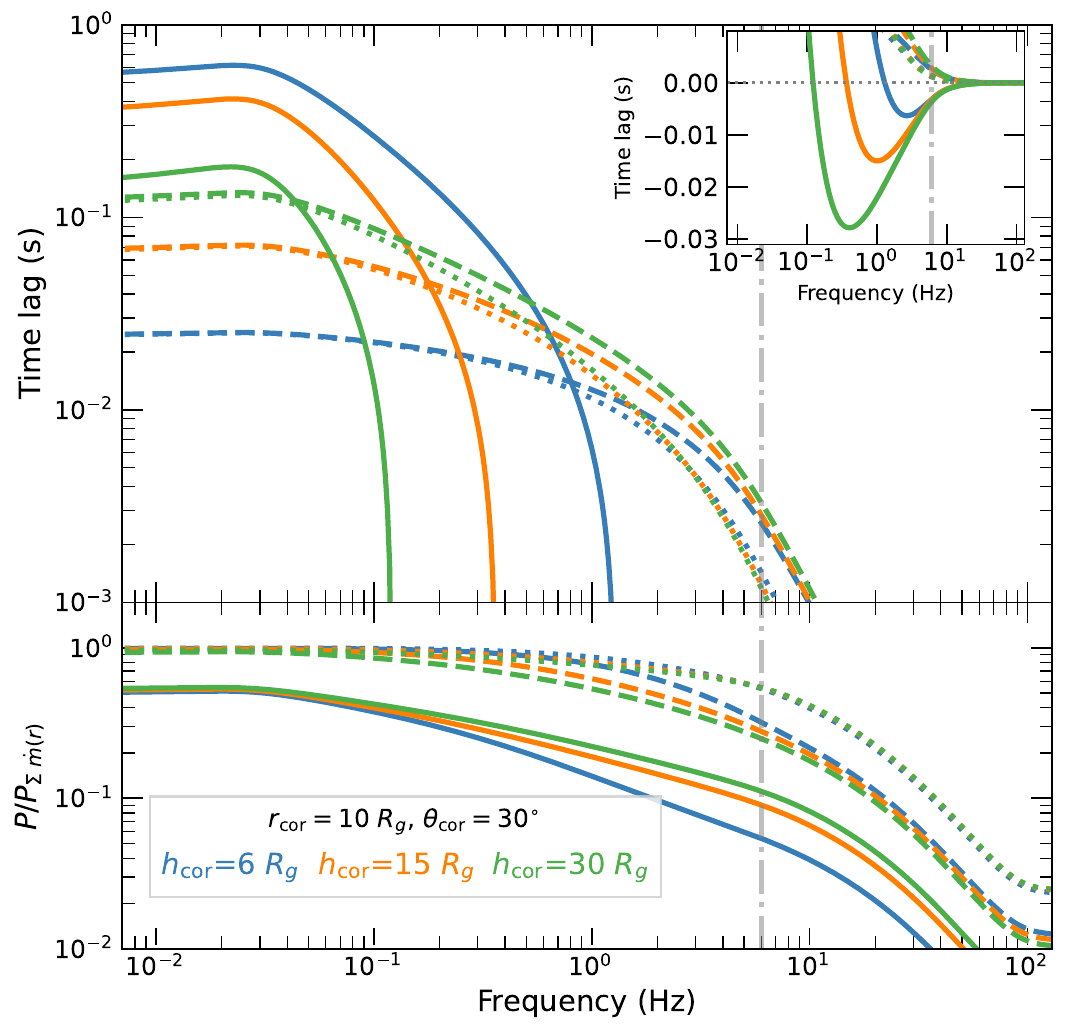} 
    \end{subfigure}
    \hfill
    \begin{subfigure}[t]{0.49\textwidth}
        \centering
        \includegraphics[width=\linewidth]{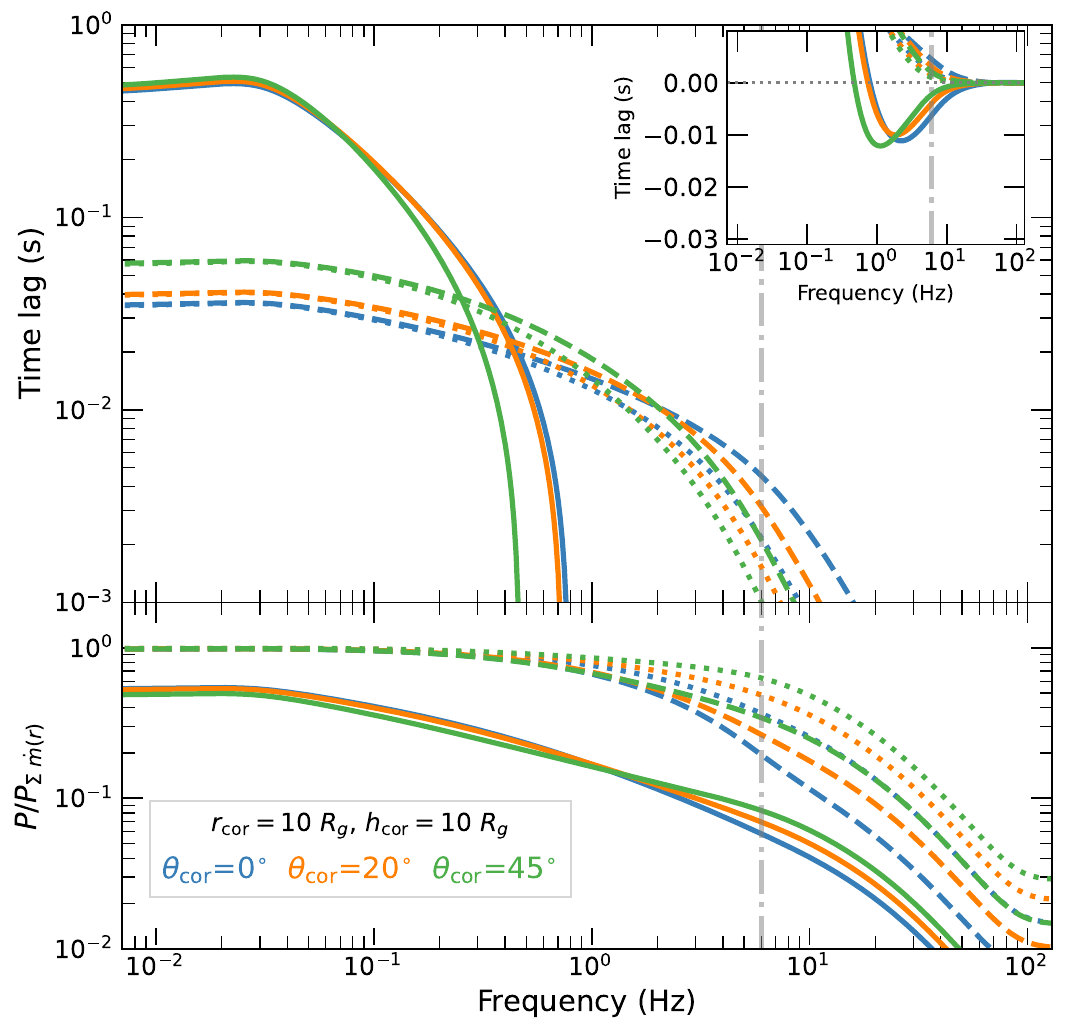} 
    \end{subfigure}
    \caption{Comparison of spectral-timing properties, corresponding by position to the different coronal geometries with impulse responses shown in Fig.~\ref{fig:irfs}. For each plot, the top part of each panel shows time lags for power-law at $E_{\rm soft}$ vs. disk (solid line), power-law at $E_{\rm med}$ vs. $E_{\rm soft}$ (dashed), power-law at $E_{\rm hard}$ vs. $E_{\rm med}$ (dotted). Positive lags indicate that the harder band lags the softer band (or the power-law lags the disk). The lower part of each panel shows the ratio of emission power spectrum  to  integrated $\dot{m}$ power spectrum for disk (solid line) and power-law flux at $E_{\rm med}$ (dashed) and $E_{\rm hard}$ (dotted). The signal peak frequencies at $r_{\rm cor}$ are shown by the vertical dot-dashed lines.} \label{fig:lagpsd}
\end{figure*}

Fig.~\ref{fig:lagpsd} shows the resulting frequency-dependent time lags and power spectra for disk and power-law emission, for the different coronal geometries with impulse responses shown in the corresponding panels in Fig.~\ref{fig:irfs}. The lags are plotted for the soft power-law flux ($3E_{\rm s}$) vs. disk luminosity and hard ($27E_{\rm s}$) vs. medium ($9E_{\rm s}$) and medium vs. soft bands. For clarity we only show power spectra for the disk and medium and hard power-law bands, since these latter bands can be observed free of disk emission in BHXRBs. We note for completeness that the soft power-law power spectrum continues the trend of greater high-frequency filtering at softer energies (see Fig.~\ref{fig:irf_explain}).

The largest changes in impulse response and corresponding lags (PL-D and different PL-PL combinations), are associated with changes in coronal radius, for the inverted cone as well as spherical coronae. Increases in coronal radius naturally shift the inner edge of disk and seed impulse responses associated with disk propagation to larger negative delays. The relative shift in disk and seed emission to larger radii produces greater suppression of high-frequency variability, which is seen in the power spectra. The resulting shift in the centroid delays increases the value of low-frequency PL-PL hard lags (as the seed impulse response is pushed to larger negative delays), but does not substantially change the PL-D hard lags at the lowest frequencies, since both disk and seed centroids are shifted in the same direction. However, the PL-D soft lags increase substantially in amplitude and shift to lower frequencies, thus decreasing the `crossing frequency' from hard to soft lags (and corresponding `cut-off' in the hard lags). This increase and shift in the soft lags is due to the change in seed centroid and because for larger coronae, reverberation dominates the disk centroid at lower signal frequencies than for coronae with smaller radii. 

The differences between spherical and inverted cone coronae can be understood in terms of their visibilities as seen from the disk, which are shown in the left panel of Fig.~\ref{fig:fd2c}. The spherical coronae generally show smaller solid angles compared to the inverted cone coronae, which have a large surface `facing' the disk. Thus spherical coronae show systematically weaker seed emission than inverted cone coronae of the same radius, as well as relatively weaker `returning' seed reverberation components for delays produced inside the corona. The latter effect leads to more negative seed centroid values for large spherical coronae compared to large inverted cone coronae, which leads to larger PL-PL hard lags and PL-D soft lags (and lower crossing-frequencies) for large spherical coronae. 

Comparing the effects of changing height and opening angle for inverted cone coronae with the same coronal radius is relatively simple and can be understood entirely in terms of the changing coronal visibility from the disk. Increases in coronal height lead to significant extensions of the propagation component of the seed impulse response to large delays, while the reverberation components do not change substantially. This pattern arises because the largest changes in visibility occur at large disk radii (see middle panel of Fig.~\ref{fig:fd2c}), which contribute the largest delays, while there is relatively minor change at small disk radii, which contribute most to reverberation. Thus the seed centroid becomes substantially more negative with increasing coronal height, PL-PL hard lags and PL-D soft lags both increase, and the PL-D crossing-frequency decreases. At the same time, there is only minor change in the power-spectra, because the large change in seed centroid delay is mainly caused by a relatively small change in seed emission at very large propagation delays.

In contrast to height changes, coronal opening angle changes produce relatively little effect on the spectral-timing properties. A change in opening angle affects the coronal visibility from the disk in a similar way over a broad range of radii (see right panel of Fig.~\ref{fig:fd2c}). Thus the seed impulse responses increase in amplitude with increasing opening angle, but the centroid delay increase is relatively small. The notable exception is in the high frequency power-law emission power spectra which show greater suppression for small opening angles. This effect is driven by the relatively large impact of coronal opening angle on the seed returning emission due to reverberation, which provides the only source of seed photons that can respond to high-frequency signals originating in the coronal part of the accretion flow.

\subsection{Evolution of spectral-timing `observables' with coronal geometry}
\label{sec:observables}
Finally, it is useful to quantify and visualise the evolution of certain spectral-timing properties which approximate the properties which can be seen in observational spectral-timing and spectral data. These are: $\langle \Gamma \rangle$, the time-averaged photon index; $\tau_{\rm max,PP}$, the maximum (low-frequency) PL-PL hard lag (fluxes for $E_{\rm hard}$ vs. $E_{\rm med}$, although the same lag is obtained for $E_{\rm med}$ vs. $E_{\rm soft}$); $\tau_{\rm min,PD}$, the minimum (or maximum negative, i.e. soft) PL-D lag (power-law flux at $E_{\rm soft}$ vs. disk luminosity); $\nu_{\rm cr, PD}$, the zero-lag crossing frequency for PL-D lags and finally $P_{\rm M}/P_{\rm H}(\nu=5\mbox{ Hz})$, the ratio of power at $E_{\rm med}$ to that at $E_{\rm hard}$, as measured at a frequency of 5~Hz.

\begin{figure*}
    \centering
    \includegraphics[width=0.8\textwidth]{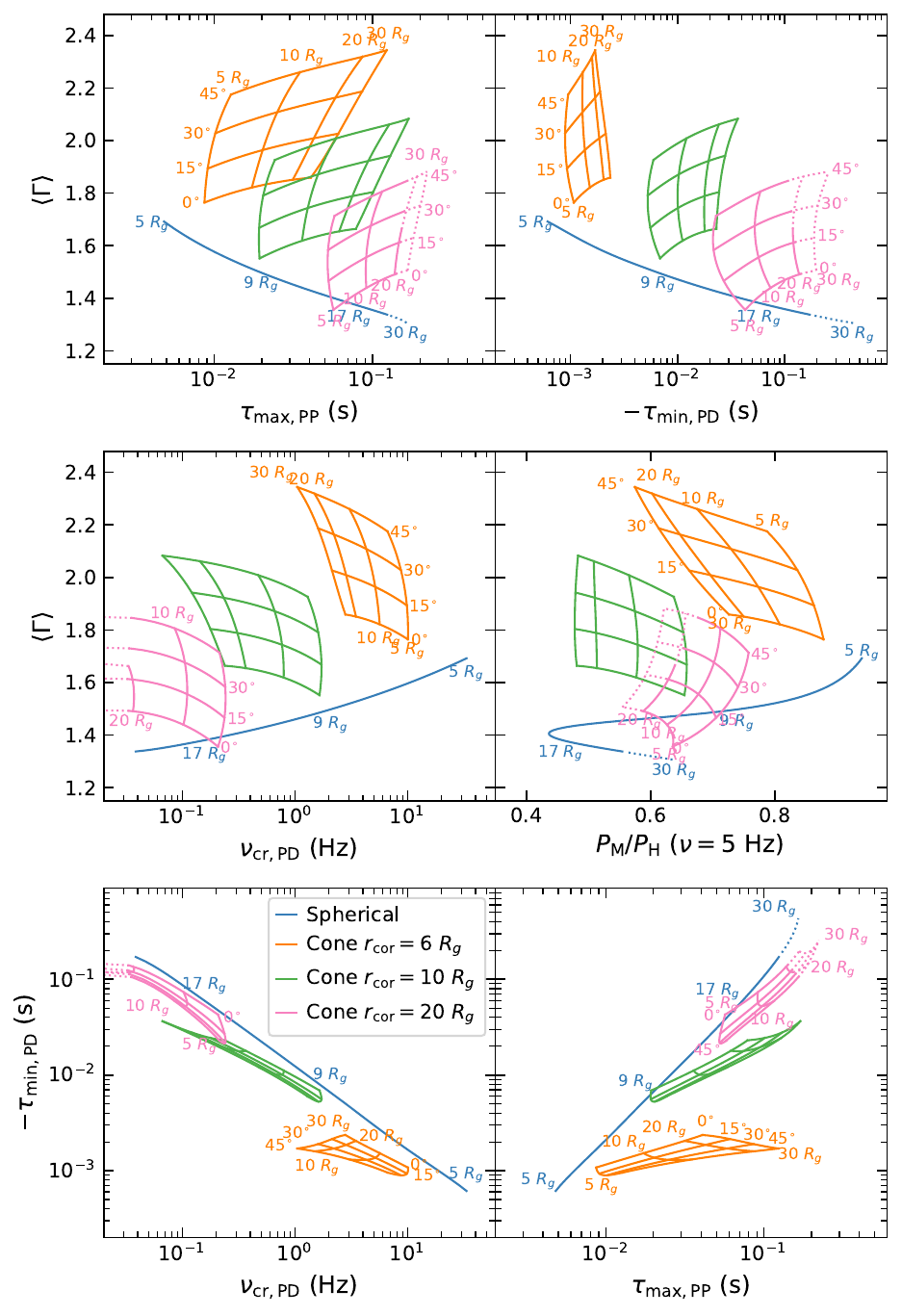} 
    \caption{Dependence of spectral-timing observable quantities on coronal geometry, for a spherical corona and inverted cone coronae (with $r_{\rm cor}=$\{6, 10, 20\}~$R_{g}$). The observable quantities are: $\langle \Gamma \rangle$, the time-averaged photon index; $\tau_{\rm max,PP}$, the maximum (low-frequency) PL-PL hard lag ($E_{\rm hard}$ vs. $E_{\rm med}$); $\tau_{\rm min,PD}$, the minimum (or maximum negative, i.e. soft) PL-D lag; $\nu_{\rm cr, PD}$, the zero-lag crossing frequency for PL-D lags and $P_{\rm M}/P_{\rm H}(\nu=5\mbox{ Hz})$, the ratio of power at $E_{\rm med}$ to that at $E_{\rm hard}$ at $\nu=5$~Hz. Coronal parameter values are labelled for the spherical and largest/smallest inverted cone coronae. See the lower-left panel for the legend for all panels.} \label{fig:observables}
\end{figure*}

Figure~\ref{fig:observables} shows these quantities obtained from the frequency-dependent lags and power-spectra calculated for the spherical and inverted cone geometries. For the spherical corona, we calculate for 36 values of $r_{\rm cor}$ spaced geometrically from 5 to 30~$R_{g}$ (corresponding to edges of the disk radial bins). For the cone geometry we calculate for each of three values of $r_{\rm cor}$ (6, 10 and 20~$R_{g}$), 160 combinations of $h_{\rm cor}$ and $\theta_{\rm cor}$ evaluated along gridlines of fixed $h_{\rm cor}=$\{5, 10, 20, 30\}~$R_{g}$ and $\theta_{\rm cor}=$\{0, 15, 30, 45\}~degrees. The coronal parameter values corresponding to set points or vertices on each grid line are labelled in the plot. For the inverted cone coronae we label for clarity only those vertices that are clearly visible and for the $r_{\rm cor}=6$~$R_{g}$ and $r_{\rm cor}=20$~$R_{g}$ cases only: the corresponding grid points for $r_{\rm cor}=10$~$R_{g}$ can be inferred from those shown for the other coronal radii.

The time-averaged photon index $\langle \Gamma \rangle$ depends on the ratio of seed to coronal heating luminosity (Eqn.~\ref{eqn:gameqn}) and thus decreases with increasing coronal radius $r_{\rm cor}$, which changes the balance of viscous dissipation from disk emission and hence seed photons, towards coronal heating. Increases in coronal opening angle increase the seed photon flux substantially, while increases in coronal height only increase the seed flux marginally (from larger radii), hence $\theta_{\rm cor}$ is strongly correlated with $\langle \Gamma \rangle$ while $h_{\rm cor}$ is only weakly correlated with $\langle \Gamma \rangle$. 

For the reasons described earlier in this subsection, the lags and crossing-frequency depend most strongly on coronal radius and height but not on opening angle. The strong dependencies on coronal radius and height result in strong (anti-)correlations between the minimum PL-D lag, crossing-frequency and maximum PL-PL lag, which can be seen in the lower two panels of Fig.~\ref{fig:observables}. The energy-dependent power-spectral evolution is more complex however, showing a maximum effect for intermediate coronal radii (between 10 and 15~$R_{g}$). This maximum effect on energy dependence of the power spectra at 5~Hz seems to correspond to the point where this frequency corresponds to the signal frequency at $r_{\rm cor}$. For signal frequencies produced within $r_{\rm cor}$, coronal heating and seed impulse responses track each other closely, since the seed emission for delays produced within the corona is produced by reverberation driven by the coronal heating power. Therefore the medium and hard band power spectra stop diverging in shape at signal frequencies produced within the corona, since they depend on an energy dependent combination of emission fractions $f_{h}$ and $f_{s}$.

\subsection{Dependence on variability and propagation time-scale parameters}
\label{sec:timescale_dependence}
For the previous analyses we have fixed the disk and coronal variability time-scale normalisation and radial scaling index to $s_{t}=1$ and $n_{t,{\rm d}}=2$ respectively (see Equation~\ref{eqn:timescales}) and the disk and coronal propagation time-scales to be equal to 100\% and 10\% of the variability time-scale respectively (corresponding to $f_\Delta\tau,{\rm d}=1$ and $f_\Delta\tau,{\rm c}=0.1$, see Equation~\ref{eqn:delays}). These values are chosen to replicate reasonably well the observed spectral-timing properties (see previous subsections and Appendix~\ref{app:diskencalc}). However, it is interesting to examine the effect of the time-scales for accretion flow variability and propagation on the observed spectral-timing properties. 
\begin{figure*}
    \includegraphics[width=0.47\textwidth]{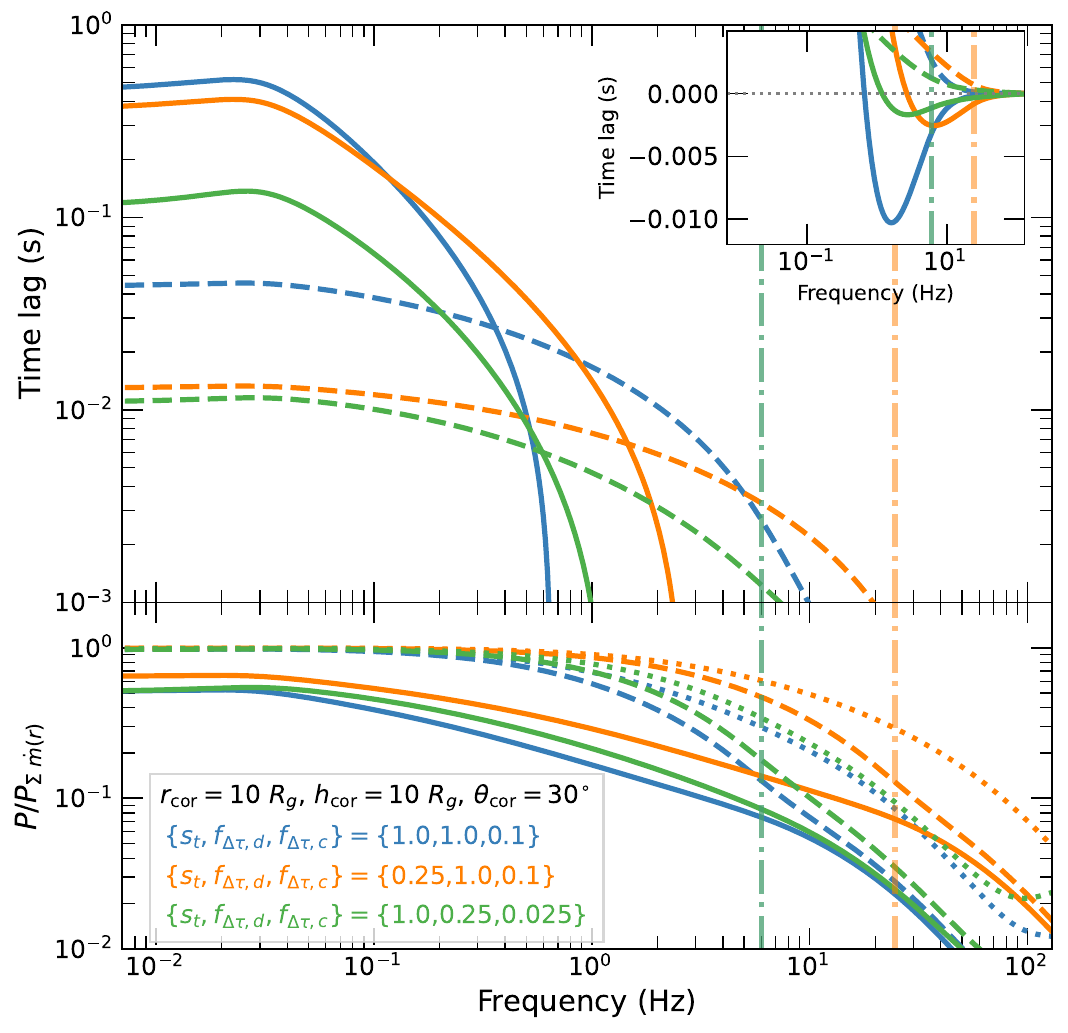}
    \includegraphics[width=0.47\textwidth]{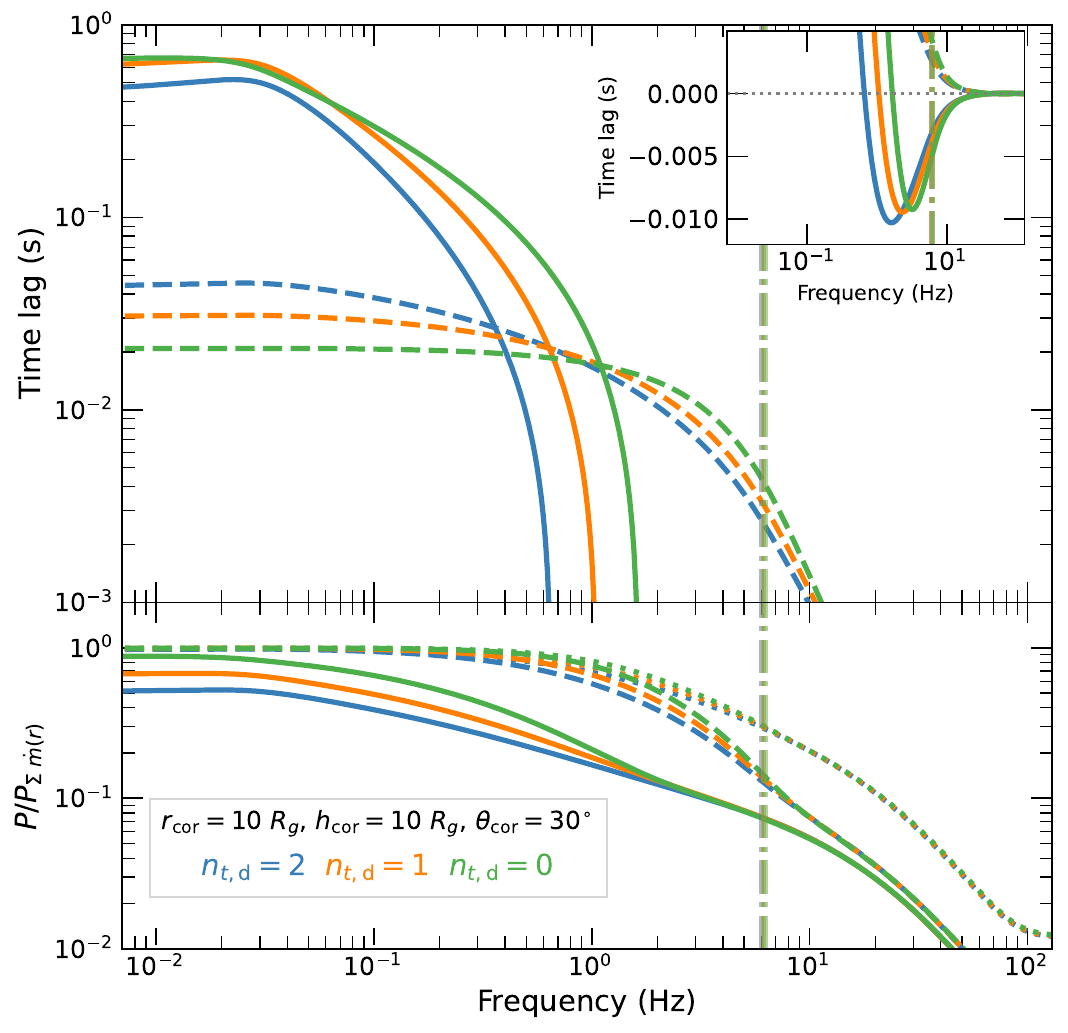}
    \caption{{\it Left panel:} Comparison of spectral-timing properties for the inverted cone corona with different scaling constants for accretion flow variability and propagation time-scales, assuming disk variability time-scale radial-scaling index $n_{t,{\rm d}}=2$ (see Equations~\ref{eqn:timescales} and \ref{eqn:delays}). The dot-dashed lines show the frequencies at $r_{\rm cor}$ for the two cases of $s_{t}$. {\it Right panel:} Comparison of spectral-timing properties for different $n_{t,{\rm d}}$. To ensure that fluctuation frequencies are identical at $r_{\rm cor}=10$~$R_{g}$ (dot-dashed lines), we set $s_{t}=\{100,10,1\}$ for $n_{t,{\rm d}}=\{0,1,2\}$.}
    \label{fig:tcompare}
\end{figure*}

The left panel of Fig.~\ref{fig:tcompare} shows the effects of changing the normalisation of disk variability time-scale and propagation time-scale relative to the default situation (shown in blue), while maintaining fixed maximum fluctuation frequency (and power-spectral break), $\nu_{\rm fluc,max}=0.03$~Hz. The main effect of reducing $s_{t}$ (orange curves) is to reduce (proportionate with $s_{t}$) the maximum PL-PL hard lags and PL-D soft lags, while increasing the zero-crossing frequency of PL-D lags and the frequencies of the roll-overs seen in the PL-PL hard lags and the power spectra. These changes occur because these spectral-timing features are produced over a range of radii which are fixed by the coronal geometry. The variability and propagation time-scales associated with these fixed radii simply scale with $s_{t}$. In contrast the PL-D lags at low frequencies are produced at radii which increase as $s_{t}$ is reduced (to maintain the fixed $\nu_{\rm fluc,max}$), so they remain relatively independent of $s_{t}$. 

In contrast, reducing the propagation time-scales relative to $s_{t}$ (green curves), by reducing $f_\Delta\tau,{\rm d}$ and $f_\Delta\tau,{\rm c}$, leads to a proportionate reduction in the amplitudes of all lags while largely maintaining the crossing-frequency of PL-D lags and the roll-over frequencies of PL-PL lags and all power spectra. Due to the requirement that propagation time-scales should be similar to the variability time-scale, it is unlikely that significant changes of propagation time-scale relative to variability time-scale will occur in real accretion flows. However, significant changes of variability time-scale could be associated with changes in $h/r$ (or possibly viscosity parameter $\alpha$) and could therefore produce evolution of spectral-timing behaviour similar to that expected from changing $r_{\rm cor}$ (e.g. see Fig.~\ref{fig:lagpsd}, upper panels), but without the concomitant changes in $\Gamma$.

The right panel of Fig.~\ref{fig:tcompare} shows the spectral-timing properties for an inverted cone corona with three different values of the radial scaling index of variability time-scale $n_{t,{\rm d}}$, assuming the same fluctuation frequencies at $r_{\rm cor}$. For standard accretion disk variability time-scales, we recall here that $n_{t,{\rm d}}=0$ corresponds to a constant aspect ratio $h/r$ of the accretion flow with radius, while $n_{t,{\rm d}}=2$ corresponds to constant $h$. The overall spectral-timing properties depend only weakly on $n_{t,{\rm d}}$, when compared to changes in coronal radius and height (e.g. see Fig.~\ref{fig:lagpsd}). This means that the model results shown here are qualitatively robust to changes in $n_{t,{\rm d}}$, but also implies that it may be difficult to determine the correct value of $n_{t,{\rm d}}$ from observations. However, the range of disk variable disk radii changes dramatically with $n_{t,{\rm d}}$, with maximum fluctuation radius $r_{{\rm fluc,max}}=\{355,86,46\}$~$R_{g}$ for $n_{t,{\rm d}}=\{0,1,2\}$, assuming a 0.03~Hz minimum fluctuation frequency for this example. These differences have strong implications for the energy-dependence of the disk impulse responses and corresponding spectral-timing properties, which we discuss in Appendix~\ref{app:diskencalc}.

\section{Discussion}
\label{sec:discussion}
We have shown that a model where mass accretion fluctuations propagate through the accretion disk to a central and relatively compact corona, can explain a wide range of the observed hard state spectral-timing properties from accreting black holes. To enable this match to observations, the effects on the coronal power-law spectrum of variable disk seed photons and coronal heating should be included, as well as reprocessing of coronal emission by the disk. The variability is produced by a plausible mechanism, namely turbulence in the accretion flow which leads to propagating mass-accretion fluctuations, with longer time-scale variations being generated at larger radii. The latter process is strongly implied by other observable properties such as the rms-flux relation \citep{Uttleyetal2005} and the broadband nature of the aperiodic variability \citep{Lyubarskii1997,IngramDone2011}, and appears naturally in (GR)MHD simulations of accretion flows \citep{HoggReynolds2016,Bollimpallietal2020}. 

Beyond these main assumptions, no special features are required for the model to explain many observed spectral-timing properties of BHXRBs along with changes in those properties, e.g. due to changing coronal geometry. We consider it a major strength of the model that it is both physically simple and consistent with the current disk plus inner corona picture of the innermost regions of accreting black holes. This disk-corona picture is supported by a wide range of observational and theoretical results, even if the nature of the corona is still to be determined \citep{Doneetal2007,IngramMotta2019,Kalemcietal2022}. We further suggest that to prevent the significant spectral-timing effects shown here, we would need to discard one or more of these assumptions which underpin our understanding of other key aspects of BHXRB spectra and variability in luminous hard states. The spectral-timing features we explain here can thus be seen as a natural consequence of the response of the disk-corona system to mass-accretion fluctuations propagating through the disk.

We now consider some of the key implications of our results for the lags predicted by Comptonisation and reverberation models, the nature of the variable accretion disk and the evolution of coronal geometry through the different spectral states. We will end with some discussion of possible improvements and extensions to the model.

\subsection{Light-travel delays due to Comptonisation and reverberation}
In Section~\ref{sec:irfmodel} we showed how a time-delay between correlated variations in coronal seed photon and heating luminosities leads naturally to energy-dependent time lags between flux variations at different power-law energies. The lag scales with the difference in centroid delays of the heating and seed impulse response (Eqn.~\ref{eqn:pllags_centroid}), so any physical scenario where there is a delay between these components should lead to a large lag in addition to that caused by other effects such as Compton scattering light-travel delays. An obvious corollary of this effect is that models which seek to explain power-law lags exclusively with Compton scattering delays (e.g. \citealt{Kazanasetal1997,Reigetal2003,Karpouzasetal2021,Bellavitaetal2022}) must include a plausible mechanism to explain why coronal heating and seed variations should be coincident in time\footnote{Note that in their Comptonisation model for quasi-periodic oscillation spectral-timing properties, \citet{Karpouzasetal2021,Bellavitaetal2022} do include separate time-variable coronal heating and seed photon oscillations, but these are set to be in phase with one another, so that Comptonisation light-travel delays dominate the lags.}. 

Our model shows that Comptonisation delays from large coronae are not needed to explain the observed PL-PL hard lags, which can be explained by much smaller coronae, and in fact seem to {\it require} relatively small coronae, or the expected lags would be much larger than observed. Such small coronae should produce Comptonisation delays $<1$~ms, which could become significant at high Fourier frequencies where propagation-induced delays become small.

Reverberation light-travel delays are not included in our model, where we assume them to be negligible compared to the propagation-related delays. This assumption is consistent with the compact coronae that are required to explain the data in our model, but such delays must be present, even if they are dominated by the expected propagation-related hard and soft lags over most of the observable frequency range. 
Fig.~\ref{fig:lagpsd} shows that the propagation-related soft lags go to zero at frequencies of tens of Hz or higher. This arises because the seed and reverberation components are directly powered by coronal heating for impulse response delays produced inside the corona, resulting in zero lag between the power-law and reverberation emission. This result suggests that `clean' light-travel reverberation signals are probably best-observed at high frequencies, $\sim100$~Hz, where any seed photon variations should be driven primarily by reverberation. Because of signal-to-noise limitations at high frequencies, such light-travel reverberation signals may be inaccessible to current X-ray spectral-timing instruments such as {\it NICER} but they will be detectable with proposed large-area missions such as the {\it enhanced X-ray Timing and Polarimetry} mission {\it eXTP} \citep{Zhangetal2019,DeRosaetal2019} and the {\it Spectroscopic Time-Resolving Observatory for Broadband Energy X-rays} ({\it STROBE-X}, \citealt{Rayetal2018}).

\subsection{The nature of the accretion disk}
The spectral-timing properties of hard state black hole X-ray binaries can be explained if the variability over a broad range of Fourier frequencies is caused by propagating mass-accretion fluctuations generated in the blackbody-emitting disk. Our results thus support with a quantitative model the scenario proposed by \citet{WilkinsonUttley2009} and \citet{Uttleyetal2011}. The speed of the propagating fluctuations is much faster than expected for a classical thin ($h\ll r$) disk however. For disk accretion, the scaling of radial infall time in our model parameterisation gives $s_{t}f _{\Delta\tau,{\rm d}}r^{n_{t,{\rm d}}}\simeq \alpha^{-1}(r/h)^{2}$, so that for our assumed $\{s_{t},f _{\Delta\tau,{\rm d}}\}=1$, $n_{t,{\rm d}}=2$ and $\alpha=0.1$, we infer $h\simeq3$ across the range of variable disk radii (e.g. $46\gtrsim r \gtrsim 10$). 

It is important to consider whether such a fast accretion disk remains significantly optically thick to Compton scattering, in order to produce the observed blackbody disk spectrum. To estimate the vertically integrated optical depth of the disk ($\tau_{z}$) we assume constant mass accretion rate $\dot{M}$ and use the continuity equation for $\dot{M}$ in terms of disk surface density $\Sigma$, radius $R=rR_{g}$ and radial velocity $v_{r}$ (e.g. see \citealt{FKR}), i.e. $\dot{M}=2\pi R \Sigma v_{r}$. From Equation~\ref{eqn:delays} we obtain $v_{r}=(s_{t}f _{\Delta\tau,{\rm d}})^{-1}r_{i}^{-(n_{t,{\rm d}}+\frac{1}{2})}c$. The vertically integrated optical depth for electron scattering of photons is $\tau_{z}=\Sigma \kappa_{\rm es}$ where $\kappa_{\rm es}$ is the electron-scattering opacity of the gas. We can therefore rewrite the mass accretion rate in terms of $\tau_{z}$, $r$, black hole mass $M$ and various constants as:
\begin{equation}
    \dot{M} = \left(\frac{2\pi GM}{\kappa_{\rm es}c}\right) \tau_{z} \left(s_{t}f _{\Delta\tau,{\rm d}}\right)^{-1} r^{\left(\frac{1}{2}-n_{t,{\rm d}}\right)}
\end{equation}
We can also write $\dot{M}$ in terms of the radiated bolometric luminosity $L$ expressed as a fraction of the Eddington luminosity $\L_{\rm Edd}=4\pi G M c/\kappa_{\rm es}$, and the radiative efficiency of the accretion flow $\eta$:
\begin{equation}
    \dot{M} = \left(\frac{4\pi GM}{\kappa_{\rm es}c}\right )\eta^{-1} \frac{L}{L_{\rm Edd}} 
\end{equation}
Substituting and rearranging the above equations yields the following expression for $\tau_{z}$:
\begin{equation}
    \tau_{z}= 2\eta^{-1} \left(\frac{L}{L_{\rm Edd}}\right) s_{t}f _{\Delta\tau,{\rm d}}\; r^{\left(n_{t,{\rm d}}-\frac{1}{2}\right)}
    \label{eqn:tauz}
\end{equation}
If we assume $\eta=0.25$ (consistent with $r_{\rm in}=2$~$R_{g}$) then for $\{s_{t},f _{\Delta\tau,{\rm d}}\}=1$, $n_{t,{\rm d}}=2$ and $L/L_{\rm Edd}=0.1$, we obtain $\tau_{z}\simeq 25$ at $r=10$~$R_{g}$, with $\tau_{z}\propto r^{3/2}$.  Note that this calculation is independent of the vertical scale-height of the disk and confirms robustly that radial infall times that are fast enough to explain the observed lags, also remain consistent with the observed blackbody emission. Therefore, the accretion disk required to explain propagation delays is both fast and optically thick. 

The inferred radial scaling of disk time-scales $n_{t,{\rm d}}=2$, means that the disk variability time-scale is proportional to the Keplerian time-scale multiplied by $r^{2}$, which implies constant $h$ vs. $r$ if the variability time-scale behaves as a standard accretion disk viscous time-scale. Such disks have not been studied using GRMHD simulations, which tend to assume constant $h/r$ by construction, since they focus on the hot inner accretion flows commonly associated with the hard state or with radiatively inefficient accretion. Besides being strongly inconsistent with the energy-dependent PL-D hard lags (see Appendix~\ref{app:diskencalc}), constant $h/r$ ($n_{t,{\rm d}}=0$) in the blackbody-emitting optically thick disk would also imply that disk vertical optical depth decreases with radius (see Equation~\ref{eqn:tauz}), which is inconsistent with the observed disk blackbody emission. If constant $h$ vs. $r$ is a better fit to the data, it would be useful to investigate whether such disks can be produced in GRMHD simulations of inner accretion flows. It is worth noting that constant $h$ corresponds to the radiation-pressure dominated inner disk of \citet{ShakuraSunyaev1973}, although such disks are generally associated with higher luminosities than observed in the hard state. Radiation-pressure dominated disks are thermally and viscously unstable \citep{LightmanEardley1974,ShakuraSunyaev1976}, but a strong vertical magnetic field could stabilise the disk \citep{Mishraetal2022}, and might be expected in the hard state.

In our model we have not explicitly considered the possibility that mass accretion fluctuations originate at discrete radii in the accretion flow, which has been suggested as a way to explain the step-like features observed in some PL-PL lag-frequency spectra and the distinct broad bumps seen in the power-spectra \citep{ArevaloUttley2006,MahmoudDone2018,MahmoudDone2018b}. Nor have we considered a discontinuity in the fluctuation time-scale, e.g. at the disk truncation radius, as an origin for these features \citep{Rapisardaetal2016}. In fact, the smoothness of the PL-D lag-frequency dependence seems to disfavour models where fluctuations originate from just a few distinct radii or where there is a sharp change in variability time-scale. E.g. the predicted bumps in the PL-D lag-frequency spectrum, and their inconsistency with the data, can be seen in examples of such models shown in \citet{Kawamuraetal2022}. One possibility is that PL-PL lags on long time-scales are associated with a different phenomenon than propagation, e.g. short-term variability of the coronal geometry (e.g. see Bollemeijer et al., submitted). Thus on certain time-scales the lags may be enhanced and we see the effect as steps in lag vs. frequency.

\subsection{Evolution of coronal geometry}
\label{sec:discuss_coronal_evolution}
In our model, the power-law vs. disk (PL-D) soft lags and the PL-PL lags between power-law bands encode information about the temporal response of seed photons to mass accretion fluctuations, which is strongly dependent on coronal geometry. Observationally, these lags are seen to evolve through the hard state in ways which are strongly correlated with the X-ray spectral shape \citep{AltamiranoMendez2015,Reigetal2018} and timing properties \citep{Pottschmidtetal2003,WangJetal2022}. We can use the lag model presented here, in particular the results in Fig.~\ref{fig:observables}, to interpret these changes in terms of the changing coronal geometry, subject to the caveat that the model presented so far remains incomplete (see Section~\ref{sec:discuss_improvements}). We also implicitly assume that spectral-timing changes are linked to coronal geometry rather than changes in accretion flow variability properties such as time-scales or maximum fluctuation radius. As shown in Section~\ref{sec:timescale_dependence}, changes in variability time-scales can mimic the effects of changes in $r_{\rm cor}$, but our model does not predict the evolution of $\Gamma$ in this case. Changes in maximum fluctuation radius may impact the maximum PL-PL hard lag, but should not significantly affect the high-frequency spectral-timing behaviour such as PL-D soft lags and crossing-frequency.

Firstly, it is notable that changes in coronal radius result in PL-PL and PL-D lag changes which are anti-correlated with the power-law photon index. This result seems to be contrary to the observed trend of increasing lags with increasing photon index \citep{Pottschmidtetal2003,Reigetal2018}. Such evolution is better explained with a simultaneous increase in both coronal height and opening angle, since neither coronal parameter on its own can cause a strong increase in both lag and photon index at the same time (see the top-left panel of Fig.~\ref{fig:observables}). One should be cautious however since the observed lags are usually calculated from the cross-spectrum averaged over a broad frequency range, for the purposes of correlating them with changes in photon index. Therefore, changes in the lag-frequency shape may also be important, but these effects are not captured in the simple analysis of the maximum lag amplitude used to make Fig.~\ref{fig:observables}. For example, above 10~Hz, the PL-PL hard lags appear to be strongly anti-correlated with coronal radius for the inverted cone geometry (see Fig.~\ref{fig:lagpsd}). Changes in coronal radius would lead to a strong positive correlation of PL-PL hard lags with photon index at those frequencies.

\citet{WangJetal2022} showed for a sample of transient BHXRBs, the strong connection of PL-D soft lags with the broadband power-spectral shape parameterised by the power-spectral hue \citep{Heiletalpowcol2015}. Our model does not predict the power-spectral properties, which depend on the assumed radial dependence of variability time-scales and the maximum radius of fluctuations (which sets the break-frequency of the power spectrum). Without speculating about the physical connection between timing and coronal geometry, we note that the timing properties also link to the spectral shape of coronal emission, with photon index increasing from the hard state to the soft-intermediate state \citep{Reigetal2018}. 

In the analysis of \citet{WangJetal2022}, PL-D soft lags are seen to first increase in amplitude from $<1$~ms to 2~ms during the hardest part of the hard state, then decrease back to $<1$~ms before increasing sharply to $>10$~ms as the source evolves from the hard-intermediate state (HIMS) to soft-intermediate state (SIMS), where broadband-noise variability is quenched \citep{Bellonietal2005,Heiletalpowcol2015}. Studying the top-right panel of Fig.~\ref{fig:observables}, we see that the decrease in PL-D soft lag during the hard state could possibly be linked to a small drop in disk truncation radius if it is associated with an increase in photon index, while the following large increase in lag during the HIMS may be explained with a combination of increasing coronal height and opening angle, which could also explain the PL-PL hard lag evolution. 

Thus, the overall picture proposed by \citet{WangJetal2021,WangJetal2022}, of increasing coronal height during the HIMS, remains applicable when considering our model. However the required increase in height is much smaller than that inferred from light-travel time lags. PL-D soft lags exceeding 10~ms can be produced by inverted cone coronae with height $\sim10$~$R_{g}$, which appear to be consistent with the relatively small coronal heights inferred from modelling of reflection features. Such coronae would also not appear as being jet-like, so they are likely to be consistent with observed X-ray polarisation signatures which imply a Comptonising region which is extended orthogonally to the radio jet axis (\citealt{Krawczynskietal2022,Veledinaetal2023}, Ingram et al. in prep.).
\begin{figure}
    \centering
    \includegraphics[width=0.45\textwidth]{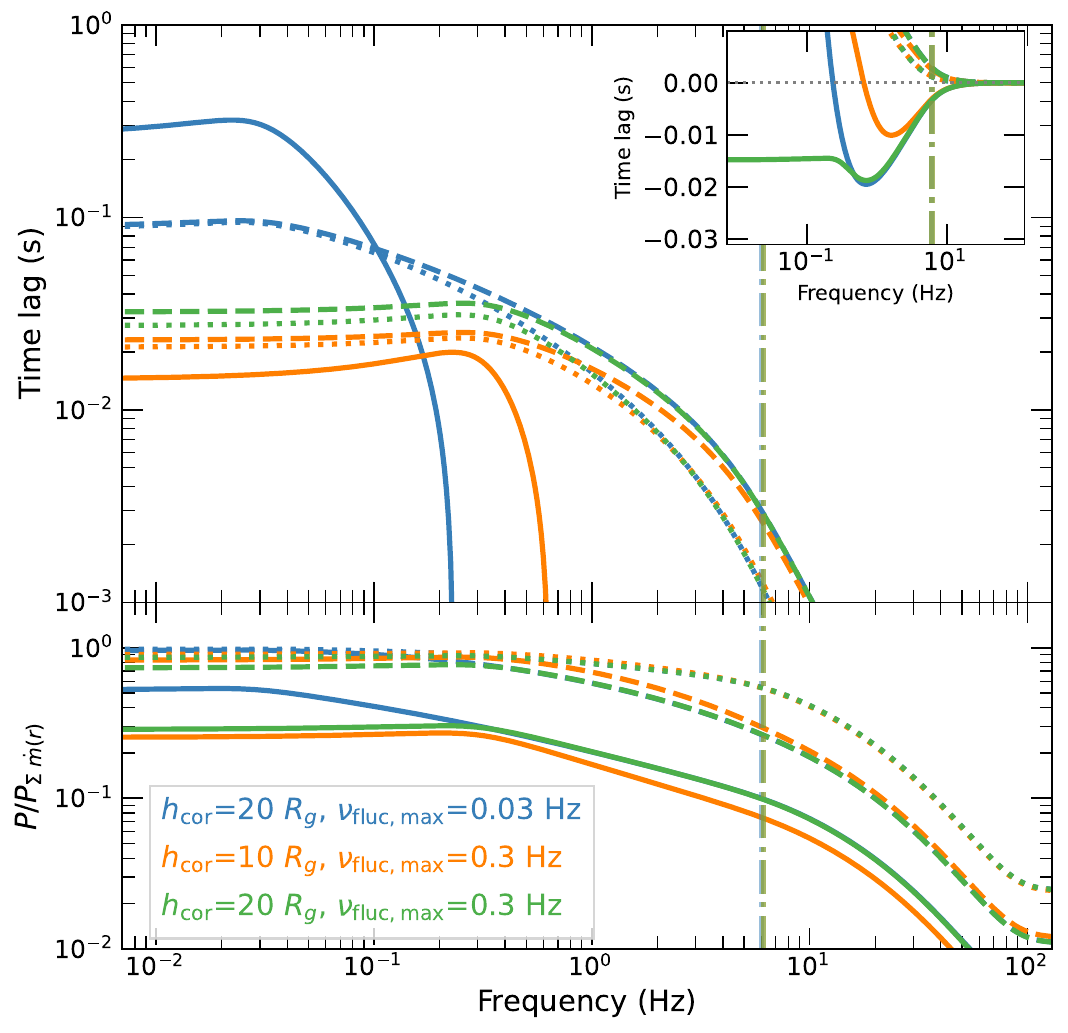}
    \caption{Comparison of spectral-timing properties for inverted cone coronae with radius $r_{\rm cor}=10$~$R_{g}$ and maximum variable disk radii of 46~$R_{g}$ and 23 $R_{g}$ producing fluctuation frequencies $\nu_{\rm fluc,max}$ of 0.03~Hz and 0.3~Hz respectively. For coronal height $h_{\rm cor}=20$~$R_{g}$ the positive PL-D hard lags seen for $\nu_{\rm fluc,max}=0.03$~Hz disappear when the maximum variable radius decreases, with only PL-D soft lags seen for $\nu_{\rm fluc,max}=0.3$~Hz. PL-D hard lags return if the corona then shrinks to $h_{\rm cor}=10$~$R_{g}$.}
    \label{fig:nuflucmax_comparison}
\end{figure}

Our model provides a natural explanation for the behaviour of BHXRBs in the HIMS which show only large PL-D soft lags across the full observed frequency range \citep{WangJetal2022}. In Fig.~\ref{fig:observables}, the lower-left panel shows that the PL-D soft lag and lag crossing-frequency (from hard to soft lags) are tightly correlated. Therefore, if the crossing-frequency is smaller than the minimum fluctuation frequency produced at the largest variable radius in the disk, i.e. the power-spectral break frequency, hard lags are never seen. For soft lags $>10$~ms this will occur for power spectral breaks $\sim 1$~Hz or smaller. We demonstrate this effect in Fig.~\ref{fig:nuflucmax_comparison}. An interesting observational example is shown by \citet{WangJetal2022} for GX~339-4 (see their Fig.~3), which shows in its intermediate state large ($>30$~ms) soft lags extending well below the break frequency at $\simeq 0.5$~Hz. \citet{Kawamuraetal2023} have interpreted the large negative lags extending to low frequencies as a consequence of seed photons with energies significantly lower than those from the inner disk, e.g. due to cyclotron emission in the hot coronal flow. However, in our model interpretation such seed photons are not required. Instead, this situation would correspond to the largest variable disk radius lying well inside the seed-producing region of the disk. This scenario might occur if the variable region of the disk shrinks, or the corona increases in height, or through some combination of these changes.

\subsection{Extensions and improvements to the model}
\label{sec:discuss_improvements}

An important caveat to our model is that it treats the seed photons as if they are mono-energetic, and furthermore does not include the inevitable spectral overlap of power-law and disk emission components which should be a consequence of the disk providing seed photons to the corona. Both these aspects can be dealt with using a full energy-dependent spectral-timing treatment to predict lag vs. energy spectra among other quantities. Such a treatment would also account for the energy-dependence of the disk lags, which we touch on briefly in Appendix~\ref{app:diskencalc}. The radial temperature-dependence of the disk leads to a strong energy-dependence of the PL-D lags which can be used to constrain the radial scaling of disk variability time-scales. For our current model it is encouraging to note that for the time-scale radial scaling index $n_{t,{\rm d}}=2$ assumed here, the lags of disk bolometric luminosity are similar to the lags at $E=2.6k_{\rm B}T_{\rm max}$ which is close to the disk blackbody flux peak (see the lower panel of Fig.~\ref{fig:disken_lags}). Thus the PL-D lag results inferred in this work for total disk luminosity are unlikely to deviate much from those obtained at the disk peak, when a full spectral treatment is used.

Our treatment of coronal geometry is currently very simple, since we consider the corona to be a solid body for the purposes of calculating the solid angle of the corona as seen from the disk and the radial dependence of coronal flux illuminating the disk. In reality, the corona is unlikely to have a sharp `edge' and one should also include optical depth effects on the fraction of seed photons intercepted by the corona, as well as obscuration of part of the direct inner disk emission from the observer. For calculating coronal illumination of the disk as well as observed power-law emission from a given direction, one should account for angular dependence of coronal emission, which will be especially relevant for non-spherical coronae such as the inverted cone considered here. A detailed treatment would require input from fully-relativistic Comptonisation models, but it is unlikely that the basic results shown here would change significantly, only the more detailed dependence of spectral-timing properties on coronal parameters. This is because the key requirement for PL-PL hard lags and PL-D soft lags to be produced by propagation, is that power-law spectral pivoting takes place in response to seed and coronal heating variations. Provided the corona is in thermal equilibrium in response to variations, spectral pivoting will occur due to photon number and luminosity conservation, independent of the details of Comptonisation models.

So far we have only considered seed photons which originate from the disk, but it is also possible that seed photons might be internally generated in the corona, e.g. via internal synchrotron emission if the corona is magnetised \citep{Veledinaetal2013}. It is easy to include such emission in our model by converting some fraction of accretion power dissipated into the corona into seed luminosity. This would have the effect of pushing the seed impulse response centroid closer to the heating centroid value, which would reduce the PL-PL hard lags and PL-D soft lags. Therefore, in the context of our model, the existence of significant internal seed photons is constrained by the observed large hard and soft lags. However, a quantitative treatment would require the spectral shape of internal seed photons to be accounted for. An interesting comparison comes from accreting neutron stars, which, assuming they possess a corona similar to black holes, are expected to produce a significant source of X-ray seed photons from the neutron star boundary layer with the accretion flow. This means that neutron stars in equivalent hard states should not show large PL-PL hard lags or PL-D soft lags, as a result of the central source of seed photons which pushes the seed impulse centroid to zero delay (Basak \& Uttley in prep.).

Finally, we note that a complete model for lags should account for the effects of strong gravity and relativistic velocities close to the black hole. Relativistic beaming due to orbital motion will significantly affect the direct disk emission seen by the observer (especially when considering energy-dependent disk lags), but not the seed impulse response seen from the central corona. However, for small disk radii, light-bending could play a significant role in enhancing the seed flux from the inner disk into the corona and at the same time amplify the coronal illumination of the inner disk while reducing illumination of larger disk radii, which would also increase the seed flux due to reverberation. Both effects would shift the seed centroid to smaller delays and thus reduce the amplitudes of both the PL-PL hard lags and the PL-D soft lags, while also reducing the PL-D hard lags due to the enhanced disk reverberation component. A natural step for the model would be to incorporate the relativistic effects on seed photons and direct disk emission with the expected reverberation light-travel delays from a fully ray-traced system e.g. following a similar approach to the {\sc reltrans} model of \citet{Ingrametal2019,Mastroserioetal2021}, albeit with an extended coronal geometry rather than a single lamppost (e.g. see \citealt{Lucchini2023}). For full consistency it would be necessary to calculate reverberation delays, coronal disk illumination and seed photon illumination of the corona using the same coronal geometry throughout.

\section{Conclusions}
\label{sec:conclusions}
We have shown that the complex and evolving pattern of Fourier-frequency and energy-dependent X-ray time lags seen in hard and hard-intermediate state BHXRBs, can be explained by propagation of mass accretion fluctuations through the disk to the inner corona. Specifically:
\begin{enumerate}
    \item Lower-frequency fluctuations are generated at larger disk radii, propagating through the disk to the corona to produce the observed large low-frequency delays of coronal power-law emission with respect to disk blackbody emission, which decrease with increasing frequency.\\
    
    \item The disk emission provides coronal seed photons, which vary in response to propagating fluctuations before the corona is heated by the same fluctuations. This leads to softening and then hardening of the power-law spectrum, producing the observed hard lags between power-law bands which increase log-linearly with energy separation and are smaller at high Fourier-frequencies.\\
    
    \item Some fraction of coronal luminosity intercepts and heats the disk to produce blackbody reverberation emission. Since variations in internal coronal heating dominate the variability of bolometric coronal luminosity while the low-energy coronal power-law variability is dominated by seed photon variations, the disk reverberation signal shows large delays compared to the soft power-law emission, which depend on the propagation time from the seed-emitting part of the disk to the corona. At high frequencies these delays dominate the power-law vs. disk lags, to produce the observed `soft' lags.\\
    
    \item The smaller fraction of disk emission originating from small radii suppresses the disk and seed photon variability at high frequencies, which leads naturally to energy-dependent suppression of high-frequency variability, with the disk and softer power-law energies showing less high-frequency variability, consistent with observations.\\
    
    \item Because of the role of seed photons in these effects, the spectral-timing properties strongly depend on coronal geometry, with coronae that are more extended (radially or horizontally) showing larger power-law vs. power-law hard lags and power-law vs. disk soft lags. However, even soft lags $>10$~ms can be explained with relatively compact coronae (heights and radii $\sim10$~$R_{g}$) because they are associated with propagation delays, not light-travel times.\\
    
    \item The observed relatively weak dependence of power-law vs. disk hard lags on disk photon energy, with lags increasing down to the power-spectral low-frequency break, imply that disk variability is generated over a relatively small range of radii (a few tens of $R_{g}$). This could be explained if the disk scale height $h$ is constant with radius.\\
\end{enumerate}
The relatively compact coronae which we infer from the lags could help to reconcile the measurements of large hard and soft lags with results obtained from spectral fitting, which imply relatively small disk truncation radii in the hard state, and X-ray polarimetry, which further implies that the corona is not significantly vertically extended. Because the model links both the power-law vs. power-law and power-law vs. disk lags with the coronal geometry, it can be used to provide a powerful independent check on spectral and polarimetric fitting results. However, a more exciting prospect would be to simultaneously fit the spectral-timing properties, X-ray spectrum and polarimetry from a BHXRB to obtain the strongest possible constraints on the disk-corona geometry and accretion flow properties. Additional strong constraints could be provided by incorporating the expected reverberation light-travel lags which should appear at frequencies $\sim 100$~Hz where propagation effects are washed out, and which will be revealed by proposed large-area X-ray spectral-timing missions such as {\it eXTP} \citep{Zhangetal2019} and {\it STROBE-X} \citep{Rayetal2018}. 

\section*{Acknowledgments}
We would like to thank Niek Bollemeijer for providing the MAXI~J1820+070 spectral-timing data shown in this paper.

\bibliographystyle{mnras} 
\bibliography{uttleyrefs}

\bsp

\appendix
\section{Coronal geometry calculations}
\label{app:geomcalc}
For simplicity we calculate the fractions of photons intercepted by the corona from the disk, and vice versa, by assuming that the corona interacts with light as a solid body. Neglecting gravitational light-bending and Doppler beaming, the fraction of disk photons from a radius $r$ that are intercepted by the corona, $f_{\rm d\rightarrow c}(r)$, depends on the solid angle subtended by the corona as seen from $r$, as well as a weighting by the cosine of the angle between the direction to the corona and the normal to the disk plane (Lambert's cosine law):
\begin{equation}
\begin{split}
f_{\rm d\rightarrow c}(r) & = \int_{\Omega_{\rm cor}} \mathbf{\hat{n}_{\rm d}}.{\rm d}\mathbf{\Omega_{\rm cor}} \bigg/ \int_{2\pi} \mathbf{\hat{n}_{\rm d}}.{\rm d}\mathbf{\Omega_{\rm sph}} \\
& = \frac{1}{\pi}\int_{\Omega_{\rm cor}} \mathbf{\hat{n}_{\rm d}}.{\rm d}\mathbf{\Omega_{\rm cor}}
\end{split}
\label{eqn:disktocor}
\end{equation}

Here $\Omega_{\rm cor}$ and $\Omega_{\rm sph}$ are the solid angles of the visible surface of the corona seen from the location on the disk at $r$ and a sphere of visibility centred on the same location, $\mathbf{\hat{n}}_{\rm d}$ is the unit vector normal to the disk plane and ${\rm d}\mathbf{\Omega} = \mathbf{\hat{p}}\,{\rm d}\Omega$ where ${\rm d}\Omega$ is the solid angle of a surface element of the sphere of visibility for a {\it visible} (i.e. not self-obscured) surface element of the corona as seen from the disk location at $r$ and $\mathbf{\hat{p}}$ is the unit vector pointing towards that disk location from the surface element. By assuming cylindrical symmetry, the calculated fraction applies to emission from the entire disk radius $r$ and not just the single disk location calculated for.
\begin{figure}
    \centering
    \includegraphics[width=0.4\textwidth]{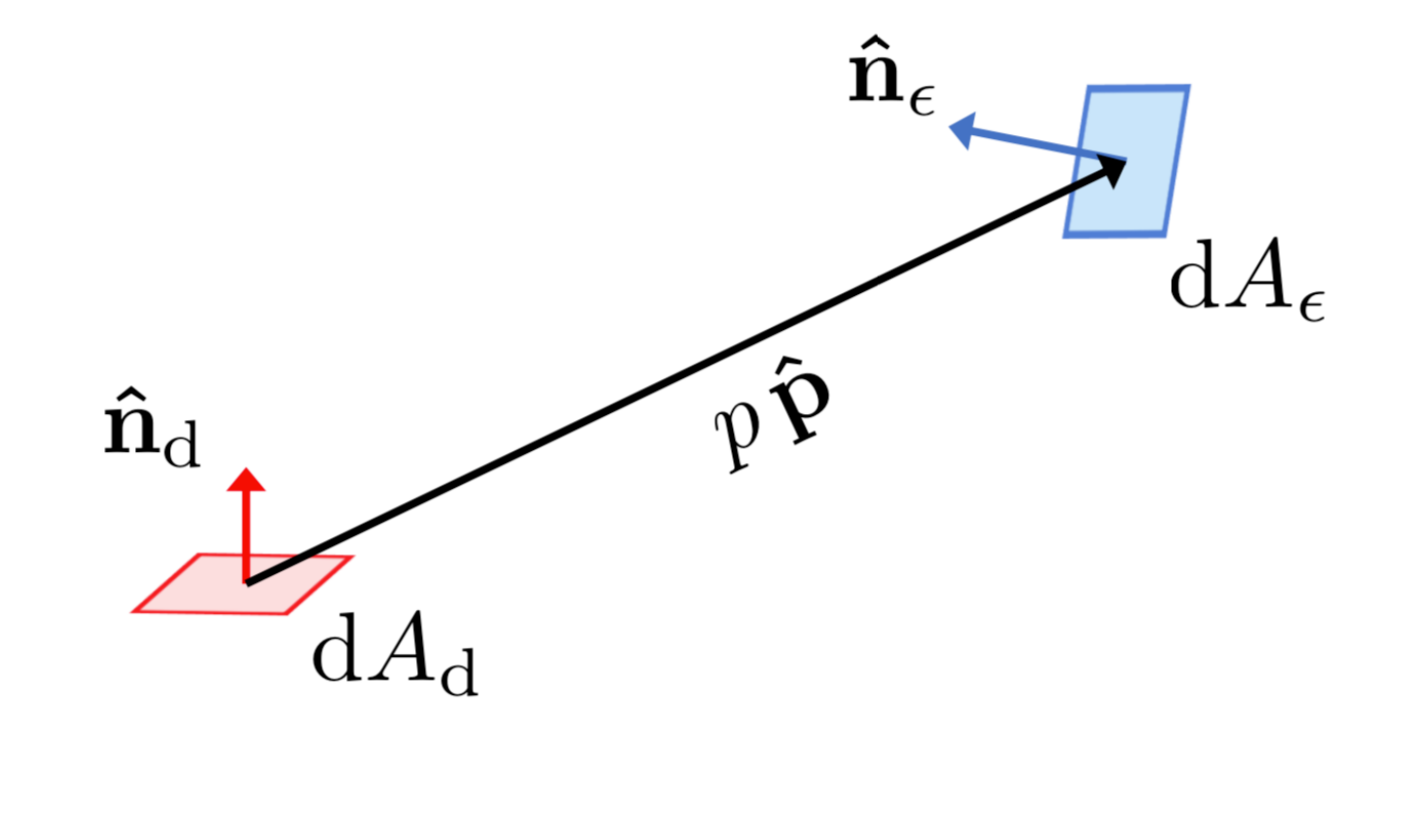}
    \caption{Setup for calculating the solid angle of a coronal surface element of area ${\rm d}A_{\epsilon}$ as seen from a disk surface element of area ${\rm d}A_{\rm d}$, and vice versa. See Appendix~\ref{app:geomcalc} for details of the calculation.}
    \label{fig:geomcalc}
\end{figure}

We illustrate the situation for calculating solid angles of surface elements in Fig.~\ref{fig:geomcalc}. The solid angle of a coronal surface element with area ${\rm d}A_{\epsilon}$ as seen from the disk at a given location, is given by ${\rm d}\Omega=\mathbf{\hat{n}}_{\epsilon}.\mathbf{\hat{p}}\,{\rm d}A_{\epsilon}/p^{2}$, where $\mathbf{\hat{n}}_{\epsilon}$ is the unit vector normal to the surface element and $p$ is the magnitude of the position vector $\mathbf{p}$ from the disk location to the surface element, i.e. $\mathbf{p}=p\,\mathbf{\hat{p}}$. Although some special cases can be solved analytically to obtain $f_{\rm d\rightarrow c}(r)$ (such as the spherical corona, see below), for most cases we must solve numerically by summing over all the coronal surface elements (denoted by the index $i$) which are visible from a location on the disk at a given radius:
\begin{equation}
\label{eqn:fd2c_sum}
    f_{\rm d\rightarrow c}(r) = \frac{1}{\pi} \sum_{\mathbf{\hat{n}}_{\epsilon,i}.\mathbf{\hat{p}}_{i}>0} \mathbf{\hat{n}}_{\rm d}.\mathbf{\hat{p}}_{i}\;\mathbf{\hat{n}}_{\epsilon,i}.\mathbf{\hat{p}}_{i}\frac{{\rm d}A_{\epsilon,i}}{p_{i}^{2}}
\end{equation}
The requirement $\mathbf{\hat{n}}_{\epsilon,i}.\mathbf{\hat{p}}_{i}>0$ ensures that the element's outward face is visible from the disk location, provided the geometry is such that no part of the corona blocks the view of the outward face. The dependence on $r$ is not explicitly shown but arises in $p_{i}$ and $\mathbf{\hat{p}}_{i}$. 

To determine the fraction of coronal emission intercepted by a unit surface area of the disk at radius $r$, $f_{\rm c\rightarrow d}(r)$, we assume that the corona has a uniform intrinsic surface brightness, with luminosity per unit surface area given by the total coronal luminosity divided by the total emitting surface area of the corona $A_{\rm cor}$. With this simplifying assumption, the fraction of total coronal luminosity, ${\rm d}f_{\epsilon}$ reaching a disk element of area ${\rm d}A_{\rm d}$ can be calculated in a similar way to the contribution of a coronal element to the fraction of disk luminosity intercepted by the corona (see Fig.~\ref{fig:geomcalc}):
\begin{equation}
{\rm d}f_{\epsilon} = \frac{1}{\pi} \mathbf{\hat{n}_{\rm \epsilon}}.\mathbf{\hat{p}} \; \mathbf{\hat{n}_{\rm d}}.\mathbf{\hat{p}}\frac{{\rm d}A_{\rm d}}{p^{2}} \frac{{\rm d}A_{\epsilon}}{A_{\rm cor}}
\end{equation}
Summing over all visible coronal elements, and multiplying by $1/{\rm d}A_{\rm d}$ to obtain the fraction of coronal emission intercepted by unit surface area of the disk, we obtain:
\begin{equation}
    f_{\rm c\rightarrow d}(r) = \frac{1}{\pi} \sum_{\mathbf{\hat{n}}_{\epsilon,i}.\mathbf{\hat{p}}_{i}>0} \mathbf{\hat{n}}_{\epsilon,i}.\mathbf{\hat{p}}_{i}\;\mathbf{\hat{n}}_{\rm d}.\mathbf{\hat{p}}_{i} \frac{{\rm d}A_{\epsilon,i}}{A_{\rm cor} p_{i}^{2}}
\end{equation}
Comparing this equation with Eqn.~\ref{eqn:fd2c_sum}, we find the simple result:
\begin{equation}
f_{\rm c\rightarrow d}(r) = \frac{f_{\rm d\rightarrow c}(r)}{A_{\rm cor}}
\label{eqn:cortodisk}
\end{equation}

For a spherical corona of radius $r_{\rm cor}$, we can solve Eqn.~\ref{eqn:disktocor} analytically to obtain:
\begin{equation}
f_{\rm d\rightarrow c}(r) = \frac{1}{\pi} \left[ \arcsin\left(\frac{r_{\rm cor}}{r}\right) - \sqrt{\left(\frac{r_{\rm cor}}{r}\right)^{2} - \left(\frac{r_{\rm cor}}{r}\right)^{4}} \right]
\end{equation}
with $f_{\rm c\rightarrow d}(r)$ obtained from Eqn.~\ref{eqn:cortodisk} assuming $A_{\rm cor}=2\pi r_{\rm cor}^{2}$ (since we only consider the coronal hemisphere on the observable side of the disk).

For the inverted cone corona modelled in this work, we solve Eqn.~\ref{eqn:fd2c_sum} numerically by splitting the corona into surface elements defined by a grid of $1000\times 1000$ bins in azimuth (from 0 to 2$\pi$) and coronal height (from 0 to $h_{\rm cor}$). To calculate $f_{\rm c\rightarrow d}(r)$ we use $A_{\rm cor}=\pi\left(r_{\rm top}^2+(r_{\rm cor}+r_{\rm top})\sqrt{(r_{\rm top}-r_{\rm cor})^2+h_{\rm cor}^2}\right)$, where $r_{\rm top}$ is the radius of the top surface of the corona ($r_{\rm top}=r_{\rm cor}+h\tan{\theta_{\rm cor}}$).

Figure~\ref{fig:fd2c} shows $f_{\rm d\rightarrow c}(r)$ calculated for the different geometries and geometric parameters, to demonstrate the effects of changing coronal radius, height and opening angle.

\section{Test of linearity with time-series simulations}
\label{app:nonlin}
We can test the validity of our use of linear impulse response functions to calculate the spectral-timing properties of our model using stochastic time-series simulations of a simplified variable accretion flow. Here we follow the approach of \citet{ArevaloUttley2006} and use the method of \cite{TimmerKoenig1995} to simulate a time series of fractional accretion rate modulation, $\dot{m}_{i}(t)$ (normalised to a mean $\langle \dot{m}_{i}(t)\rangle=1$) for each annulus of the disk corresponding to radius $r_{i}$, using the Lorentzian parameters associated with each radius to determine the $\dot{m}_{i}(t)$ power spectrum. The physical accretion rate time-series $\dot{M}_{i}(t)$ for a given radius is produced by multiplying together the $\dot{m}(t)$ generated at that radius with the physical time-series generated for the preceding (outer) radius and shifting the resulting signal by the propagation delay across the radial bin, $\Delta\tau_{i}$:
\begin{equation}
    \dot{M}_{i}(t+\Delta\tau_{i}) = \dot{m}_{i}(t)\dot{M}_{i+1}(t)
\end{equation}
By repeating this calculation for successive inwards radii, we can generate an accretion fluctuation time series for each radius, from which we can determine the dissipation power time series for each radius, by multiplying by $f_{\rm diss}(r_{i}$, calculated from Eqn.~\ref{eqn:f_diss}). This can in turn be used to generate the disk dissipation and coronal heating time series associated with each radius, the seed contribution from disk dissipation (using $f_{\rm{d}\rightarrow\rm{c}}(r_{i}$) and from summing their totals over all radii we can determine the total heating, disk and seed light curves, including reverberation contributions to the disk and seed contributions (by applying $f_{\rm{c}\rightarrow\rm{d}}(r_{i})$. Finally, we can use the generated time-series directly in Eqn.~\ref{eqn:gameqn} to determine a time-series for photon index $\Gamma$, and use Eqn.~\ref{eqn:plfluxeqn} to determine the power-law flux at a given energy.

This approach provides a numerical {\it non-linear} time-series method for calculating spectral-timing properties using our lags model. The method is non-linear for two reasons. Firstly the generated accretion rate time-series are formally non-linear, incorporating the rms-flux relation and log-normal flux distributions that are observed in the X-ray variability of accreting black holes \citep{Uttleyetal2005}. Secondly, the power-law flux variations are generated from the full, non-linear versions of Equations~\ref{eqn:gameqn} and \ref{eqn:plfluxeqn} rather than the linearised versions that are used to determine the impulse responses of the power-law flux.

\begin{figure}
    \includegraphics[width=0.47\textwidth]{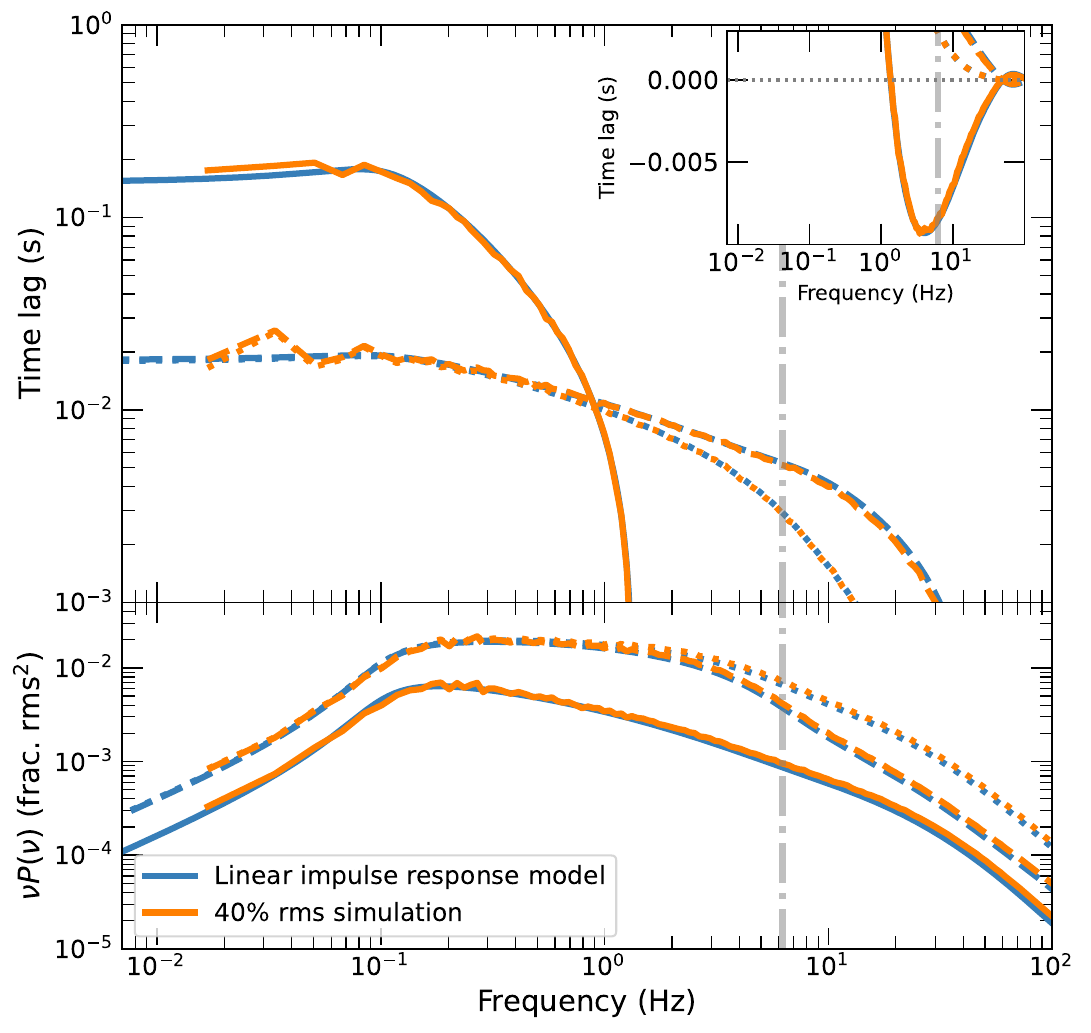}
    \caption{Comparison of frequency-dependent spectral-timing properties for numerical simulations of our lags model using a simplified accretion flow with 40~per~cent total rms variability of accretion rate, versus the results obtained from linear impulse response calculations, assuming a spherical corona with $r_{\rm cor}=10$~$R_{g}$. {\it Top panel}: time lags for power-law at $E_{\rm soft}$ vs. disk (solid line), power-law at $E_{\rm med}$ vs. $E_{\rm soft}$ (dashed), power-law at $E_{\rm hard}$ vs. $E_{\rm med}$ (dotted). Positive lags indicate that the harder band lags the softer band (or the power-law lags the disk). {\it Lower panel}: power spectra for disk (solid line) and power-law flux at $E_{\rm med}$ (dashed) and $E_{\rm hard}$ (dotted). The signal peak frequency at $r_{\rm cor}$ is shown by the vertical dot-dashed grey line.}
    \label{fig:simcompare}
\end{figure}

Figure~\ref{fig:simcompare} shows the spectral-timing results of our numerical simulation assuming a spherical coronal geometry ($r_{\rm cor}=10$~$r_{g}$) and with disk and coronal variability and propagation time-scales as described in Section~\ref{sec:num_imp}, except that we assumed a fluctuation frequency at the maximum variable radius $\nu_{\rm fluc,max}=0.1$~Hz, to reduce the required light curve lengths and thus save computation time. The rms amplitude of accretion variability is set to be 40~per~cent, consistent with variations often observed in the hard state. For the simulation, 350 sets of light curves, each consisting of $2^{22}$ 14.1~$\mu$s time bins ($\equiv 59.3$~s per light curve) were simulated and used to calculate the spectral timing products, which are further binned geometrically in frequency, with geometric binning factor 1.05. For comparison we show the corresponding results for the impulse response approach that is used in this work. The only significant deviations that can be seen at low frequencies are due to the intrinsic noise variability of the simulated time-series, which lead to fluctuations in the time-lags and the signal power (and hence the power spectra of emission). At higher frequencies this noise in the simulated data is smoothed over, but small shifts in the power spectra can be seen with power enhanced by up to $\sim20$\% at high frequencies in the simulated data. Only very small deviations are seen in the lags however. The increase in the high-frequency power from the simulated data is a known effect of the multiplicative variability process used to generate the light curves \citep{Uttleyetal2005}. However, like the lags, power-spectral ratios between different bands are not changed significantly. We conclude that even for relatively large amplitudes of variability, our linear impulse response model is an excellent approximation to the full non-linear model.

\section{Disk energy-dependent lags as a constraint on the radial scaling of disk variability time-scales}
\label{app:diskencalc}
In this work we calculate spectral-timing properties using the bolometric disk luminosity, mainly for simplicity and generalisability. A complete energy-dependent treatment of the lags involving the disk emission also requires consideration of the overlap of the disk and coronal emission in energy, including a more realistic seed photon spectrum. We leave this detailed treatment to a future work. However it is useful to note an important constraint that the energy-dependence of the disk impulse response places on the radial scaling index of the disk variability time-scale, $n_{t,{\rm d}}$ (see Equations~\ref{eqn:timescales} and \ref{eqn:delays} and Section~\ref{sec:timescale_dependence}). 

Accretion disk emission is expected to take the form of a multi-temperature blackbody, with radial temperature due to viscous dissipation alone scaling as $k_{\rm B}T\propto r^{-3/4}$ for $r\gg r_{\rm in}$ \citep{Makishimaetal1986}. Since the disk radius for a fixed fluctuation signal frequency $\nu_{\rm fluc}$  scales as $\nu_{\rm fluc}^{-1/(n_{t,{\rm d}}+3/2)}$, a given range of fluctuation frequencies should cover a greater range of disk temperature for smaller $n_{t,{\rm d}}$ and thus show a stronger dependence of the disk spectral-timing properties on observed energy. In particular we expect to see a maximum PL-D hard lag which depends on photon energy and $n_{t,{\rm d}}$, as the exponential cut-off of the blackbody spectrum limits the emission at a given energy to radii above a certain temperature, and hence limits the extent of the impulse response towards negative delays.

\begin{figure}
    \centering
    \begin{subfigure}[t]{0.45\textwidth}
        \centering
        \includegraphics[width=\linewidth]{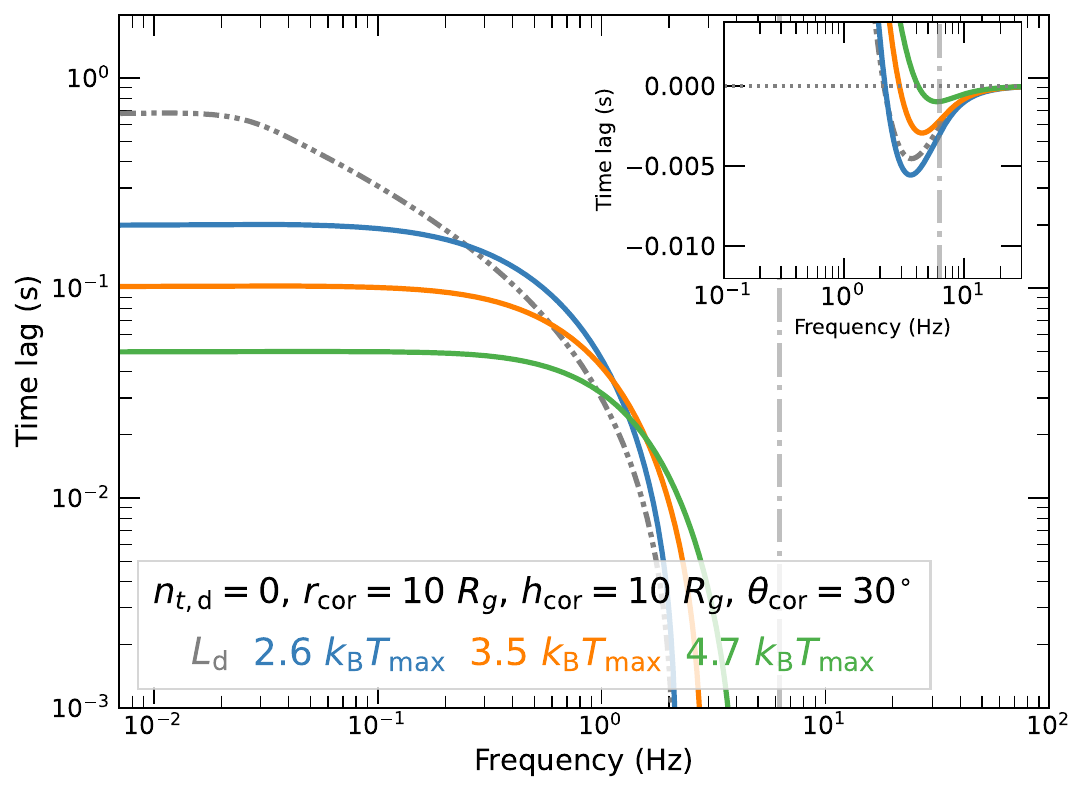} 
    \end{subfigure}
    \begin{subfigure}[t]{0.45\textwidth}
        \centering
        \includegraphics[width=\linewidth]{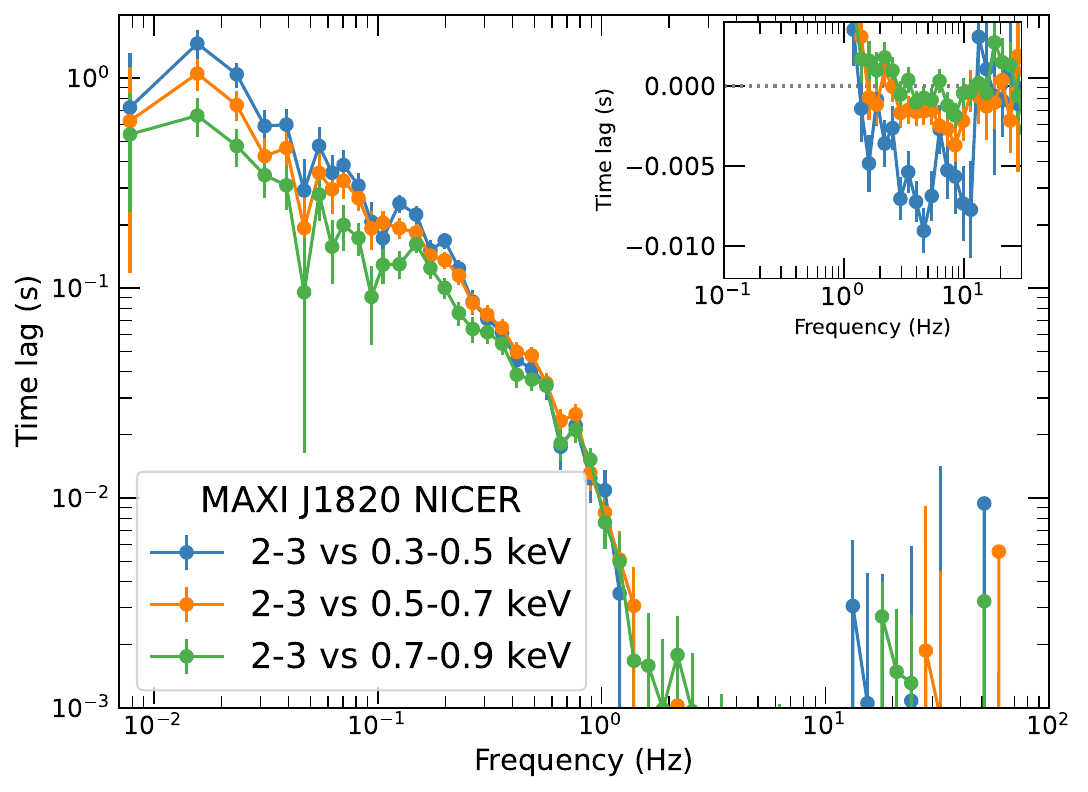} 
    \end{subfigure}
    \begin{subfigure}[t]{0.45\textwidth}
        \centering
        \includegraphics[width=\linewidth]{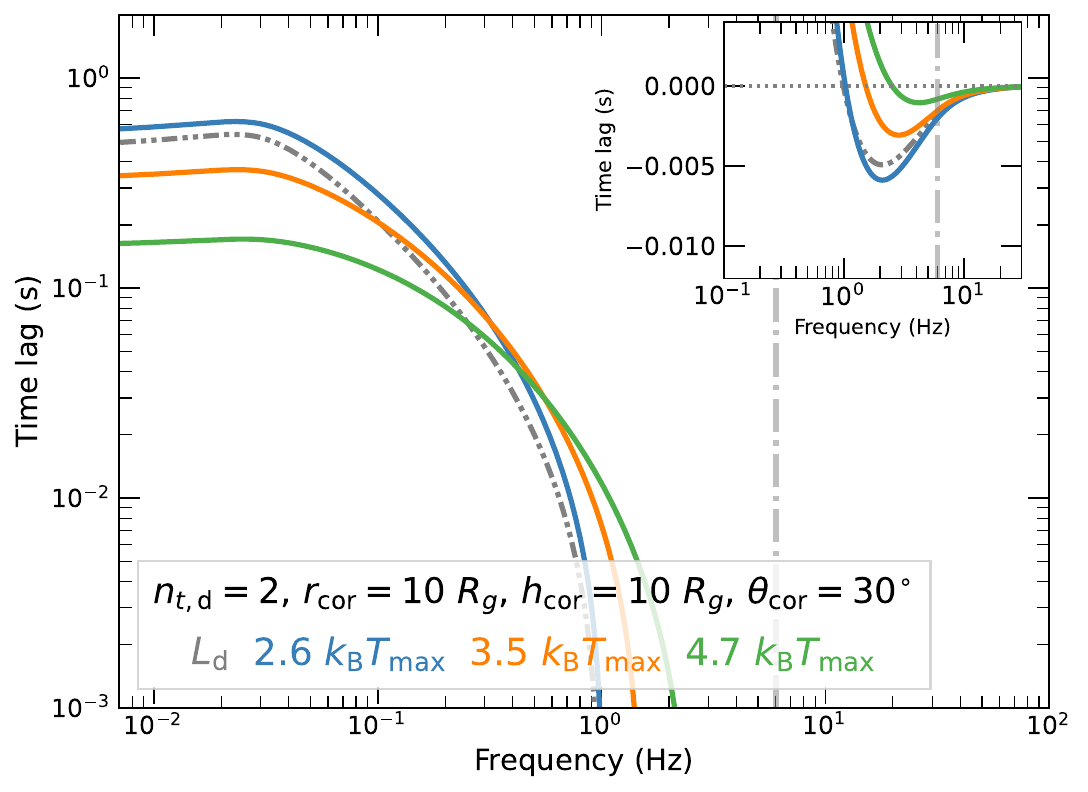} 
    \end{subfigure}    
    \caption{Energy-dependence of power-law vs. disk lags for an inverted cone corona with $n_{t,{\rm d}}=0$ and $2$ ({\it upper} and {\it lower} panels) and for comparison ({\it centre} panel) the MAXI~J1820+070 observed lags for three different energy bands covering the accretion disk emission, obtained from NICER ObsID 1200120104 (see Section~\ref{sec:intro} for further details).} \label{fig:disken_lags}
\end{figure}

To determine how the energy-dependence of disk lags depends on $n_{t,{\rm d}}$, we calculate the energy-dependent disk impulse responses using the following approach. First we use the radially-dependent disk dissipation and reverberation quantities defined in Section~\ref{sec:disk_irfs} to define a time-averaged total intensity at the radius $r_{j}$:
\begin{equation}
    I(r_{j}) = \frac{f_{\rm diss}(r_{j})}{2\pi r_{j}} + \frac{f_{\rm therm}f_{\rm c\rightarrow d}(r_{j}) \left(\langle L_{\rm s, diss}\rangle + \langle L_{\rm h}\rangle\right)}{1-f_{\rm return}}
\end{equation}
We then assume that the total intensity is proportional to the blackbody intensity at that radius, which further assumes that coronal emission reprocessed by the disk contributes to and is reprocessed as a blackbody at the local disk temperature. This latter assumption may be a reasonable approximation to the reprocessed spectrum at high disk densities (e.g. see figures and discussion in \citealt{RossFabian2007,Garciaetal2016}). Thus we can define the radial disk temperature $k_{\rm B}T(r_{j})$ relative to the maximum radial disk temperature $k_{\rm B}T_{\rm max}$ (which is observable) as:
\begin{equation}
\label{eqn:kTrad}
    k_{\rm B}T(r_{j}) = k_{\rm B}T_{\rm max} \left(\frac{I(r_{j})}{\max\; I(r_{j})}\right)^{\frac{1}{4}}
\end{equation}
For a given blackbody temperature $k_{\rm B}T$ we can calculate what fraction of the blackbody photons are emitted within an energy range $E_{1}$--$E_{2}$, which for the constant component of disk emission is:
\begin{equation}
\label{eqn:bandfrac_const}
    f_{\rm bb,const}=\frac{1}{2\zeta(3)(k_{\rm B}T)^{3}} \int^{E_{2}}_{E_{1}} \frac{E^{2}}{\left(\exp(E/k_{\rm B}T)-1\right)}{\rm d}E
\end{equation}
where the term inside the integral is the blackbody photon rate spectrum $N_{\rm bb}(E)$ and $\zeta(3)=1.2020569\dots$ is the Riemann zeta function for argument 3, also known as Ap\'{e}ry's constant. Note that we calculate the fraction of total photon rate due to photons between $E_{1}$, $E_{2}$ rather than the fraction of luminosity in this range, because this is closer to the observed quantity which is usually a photon count rate.

For the disk impulse response used to calculate the time-variation of the emission, we cannot use the time-averaged blackbody spectrum, because although the blackbody bolometric intensity scales linearly with e.g. mass accretion rate or illuminating coronal flux, the blackbody spectrum changes shape in response to the corresponding temperature change.   We must therefore obtain the linear change in blackbody photon flux spectrum $\delta N_{\rm bb}(E)\approx \frac{{\rm d}N_{\rm bb}(E)}{{\rm d}(k_{\rm B}T)}\delta (k_{\rm B}T)$ in response to a small temperature change $\delta (k_{\rm B}T)=\frac{1}{4}k_{\rm B}T\frac{\delta I}{I}$, where $\frac{\delta I}{I}$ is the relative change in bolometric intensity\footnote{See \citet{vanParadijsLewin1986} for a related calculation of the change in blackbody flux spectrum seen in a thermonuclear X-ray burst.}. The fraction of photons from this variable component contained within $E_{1}$--$E_{2}$ is then:

\begin{equation}
\label{eqn:bandfrac_var}
    f_{\rm bb,var}=\frac{1}{6\zeta(3)(k_{\rm B}T)^{4}} \int^{E_{2}}_{E_{1}} \frac{E^{3}\exp(E/k_{\rm B}T)}{\left(\exp(E/k_{\rm B}T)-1\right)^{2}}{\rm d}E
\end{equation}

With the above equation we can write the impulse response for the disk photon rate at radius $r_{j}$, with total time-delay $\tau_{j}$ and propagation delay across the radial bin ${\rm d}\tau_{j}$, as:

\begin{equation}
\label{eqn:irf_diskendisp}
g_{\rm d, diss}(\tau_{j},E_{1},E_{2}){\rm d}\tau_{j} = \frac{C_{\rm bb,var} f_{\rm bb,var}(r_{j},E_{1},E_{2})}{k_{\rm B}T(r_{j})} g_{\rm d, diss}(\tau_{j}){\rm d}\tau_{j}
\end{equation}

where $C_{\rm bb,var}=\frac{45\zeta(3)}{2\pi^{4}}$ and $C_{\rm bb,var}/k_{\rm B}T(r_{j})$ corrects the bolometric luminosity of the impulse response to a scaled total photon rate, which is then converted by $f_{\rm bb,var}$ to the rate between $E_{1}$, $E_{2}$. 

Using a similar approach we also obtain the reverberation impulse response, which must be weighted and summed over all disk radii for each impulse response bin of the illuminating coronal emission.

\begin{multline}
\label{eqn:irf_diskenrev}
g_{\rm d, rev}(\tau_{j},E_{1},E_{2}){\rm d}\tau_{j} = \frac{C_{\rm bb,var} f_{\rm therm} \left(g_{\rm s, diss}(\tau_{j}) + g_{\rm h}(\tau_{j})\right)}{1-f_{\rm return}} \\
\times \sum\limits_{i} \frac{f_{\rm bb,var}(r_{i},E_{1},E_{2}) A_{i} f_{\rm c\rightarrow d}(r_{i})\left(1-f_{\rm d\rightarrow c}(r_{i})\right)}{k_{\rm B}T(r_{i})}
\end{multline}

where $A_{i}$ is the area of the disk annulus for radial bin $r_{i}$, which is required because $f_{\rm c\rightarrow d}$ is the fraction of coronal emission intercepted by the disk {\it per unit disk area}. By adding together the above impulse responses we can obtain the impulse response of disk photon rate between $E_{1}$, $E_{2}$.

In Fig.~\ref{fig:disken_lags} we show in the upper and lower panels the calculated PL-D lags for disk energies (relative to the maximum temperature) $2.6 k_{\rm B}T_{\rm max}$, $3.5 k_{\rm B}T_{\rm max}$ and $4.7k_{\rm B}T_{\rm max}$, as well as the bolometric luminosity $L_{\rm d}$ used elsewhere in this paper. Lags are calculated relative to a power-law energy $4.9E_{\rm s}$. These energies all correspond to the median detected photon energies in each energy band used for the observational PL-D lag measurements shown in the centre panel, assuming a peak disk temperature $k_{\rm B}T_{\rm max}=0.17$~keV for this observation (N. Bollemeijer, private communication) and mono-energetic seed photons with energy $2.82k_{\rm B}T_{\rm max}$. The lags are calculated for two values of disk time-scale radial scaling index, $n_{t,{\rm d}}=0$ and $2$ (corresponding respectively to constant aspect ratio $h/r$ and constant scale height $h$), assuming an inverted cone corona with $r_{\rm cor}=10$~$R_{g}$, $h_{\rm cor}=10$~$R_{g}$ and $\theta_{\rm cor}=30^{\circ}$ and parameters as given in Section~\ref{sec:num_imp} but with $s_{t}=100$ and $s_{t}=1$ respectively, such that the disk variability time-scales are identical at $r_{\rm cor}=10$~$R_{g}$.

The $n_{t,{\rm d}}=0$ case shows a distinct energy-dependent low-frequency flattening which occurs at substantially higher frequencies than the minimum signal frequency produced in the disk. This is because, for this radial dependence of time-scales, at these energies only the inner variable part of the disk is emitting, and the extent of the variable emitting region (and hence the impulse response) is also energy-dependent. For $n_{t,{\rm d}}=2$ however, the lags continue to rise down to the lowest variable signal frequency in the disk, since the varying region is more compact for this radial scaling\footnote{The fluctuation frequency at the largest variable radius, $\nu_{{\rm fluc,max}}=0.03$~Hz used here, corresponds to $r_{\rm fluc,max}=46$~$R_{g}$ for $n_{t,{\rm d}}=2$, versus $r_{\rm fluc,max}=355$~$R_{g}$ for $n_{t,{\rm d}}=0$.}. Although some differences remain, the MAXI~J1820+070 data appears to be much more consistent with the lag behaviour for $n_{t,{\rm d}}=2$, even though the model has not been fitted to match the data in any way.

\label{lastpage}

\end{document}